\documentclass[fleqn,usenatbib]{mnras}

\usepackage[T1]{fontenc}
\usepackage{ae,aecompl}

\usepackage{graphicx}	
\usepackage{sidecap}
\usepackage{amsmath}	
\usepackage{amssymb}	
\usepackage{pdflscape}

\newcommand{\msunyr}{\ensuremath{\mathit{M}_{\odot}{\rm \ yr}^{-1}}}   
\newcommand{\kms}{\ensuremath{{\rm km \ s^{-1}}}}                   

\newcommand{\mdot}{\ensuremath{\dot{M}}}                             
\newcommand{\vinf}{\ensuremath{v_{\infty}}}                          

\newcommand{\ecar}{\ensuremath{\eta ~\mathrm{Car}}}

\def\chandra{{\it Chandra}}
\def\rxte{{\it RXTE}}
\def\hst{{\it HST}}
\def\iue{{\it IUE}}
\def\stis{{\it STIS}}

\def\fuse{{\it FUSE}} 

\def\foc{{\it FOC}}
\def\fos{{\it FOS}}
\def\stis{{\it STIS}}
\def\wfpc2{{\it WFPC2}}

\title[Fossil Winds of Eta Carinae]{The Fossil Wind Structures of Eta Carinae: \\ Changes across one 5.54-year Cycle}

\author[T. R. Gull et al.]{Theodore R. Gull,$^{1}$
\thanks{E-mail: Theodore.R.Gull@nasa.gov}
Thomas I. Madura,$^{1,2}$
Mairan Teodoro,$^{1,2}$\newauthor
Nicola Clementel,$^{3}$ 
Michael Corcoran,$^{1,2}$ 
Augusto Damineli,$^{4}$
Jose H. Groh,$^{5}$\newauthor
Kenji Hamaguchi,$^{1,6}$
D. John Hillier,$^{7}$
Anthony F. J.  Moffat,$^{8}$\newauthor
Noel D. Richardson,$^{9}$
Gerd Weigelt,$^{10}$
\newauthor
Don Lindler,$^{1,11}$
Keith Feggans$^{11,12}$
\\
$^{1}$Code 660, Astrophysics Science Division,
Goddard Space Flight Center,
Greenbelt, MD 20771 USA\\
$^{2}$Universities Space Research Association, 7178 Columbia Gateway Drive, Columbia, MD 20146  USA\\
$^{3}$South African Astronomical Observatory, P. O. Box 9, Observatory 7935, South Africa\\
$^{4}$Instituto de Astronomia, Geofisica e Ciencias Atmosfericas, Universidade de Sao Paulo, Rua do Matao 1226,\\ Cidade Universitaria, Sao Paulo 05508-900, Brazil\\
$^{5}$School of Physics, Trinity College Dublin, The University of Dublin, Dublin 2, Ireland\\
$^{6}$Department of Physics, University of Maryland, Baltimore County, 1000 Hilltop Circle, Baltimore, MD 21250, USA\\
$^{7}$Department of Physics and Astronomy and Pittsburgh Particle physics, Astrophysics, and Cosmology Center (PITT PACC),\\ University of Pittsburgh, 3941 OHara Street, Pittsburgh, PA 15260, USA\\
$^{8}$Departement de physique and Centre de Recherche en Astrophysique du Quebec, Universite de Montreal, CP 6128 Succ. A.,\\ Centre-Ville, Montreal, Quebec H3C 3J7, Canada\\
$^{9}$Ritter Observatory, Department of Physics and Astronomy, The University of Toledo, Toledo, OH 43606-3390, USA\\
$^{10}$Max-Planck-Institut fur Radioastronomie, Auf dem Hugel 69, 53121, Bonn, Germany\\
$^{11}$Sigma Space Corporation, 4600 Forbes Blvd., Lanham, MD 20706 USA\\
$^{12}$Code 670, Heliophysics Science Division, Goddard Space Flight Center, Greenbelt, MD 20771 USA 
}

\date{Accepted XXX. Received YYY; in original form ZZZ}

\pubyear{2016}

\begin{document}
\label{firstpage}
\pagerange{\pageref{firstpage}--\pageref{lastpage}}
\maketitle

\begin{abstract}
Eta Carinae, the closest, active, massive binary containing a highly unstable LBV, exhibits expanding, compressed wind shells, seen in emission, that are spatially and spectrally resolved by  \hst/\stis. Starting in  June 2009, these structures were mapped across its 5.54-year, highly elliptical,  binary orbit to follow temporal changes in  the light of [Fe~III] 4659\AA\ and [Fe~II] 4815\AA. The  emissions trace portions of fossil wind-shells, that were formed by wind-wind interactions across each cycle.  Over the high-ionization state, dense arcs, photo-ionized by far ultraviolet radiation from the hot secondary, are seen in [Fe~III]. Other arcs, ionized by mid-ultraviolet radiation from the primary star, are seen in [Fe~II]. The [Fe~III] structures tend to be interior to [Fe~II] structures that trace  extensive, less disturbed primary wind. During the brief periastron passage when the secondary plunges deep into the primary's extremely dense wind,  on the far side of primary star, high-ionization [Fe~III] structures fade and reappear in [Fe~II].  
Multiple fossil wind-structures were traced across the 5.7-year monitoring interval. The strong similarity  of the expanding [Fe~II] shells  suggests that the wind and photo-ionization properties of the massive binary have not changed substantially from one orbit to the next over the past several orbital cycles.
These observations trace  structures that can be used to test three-dimensional hydrodynamical and radiative-transfer models of  massive, interacting winds. They also provide a baseline for following future changes in  \ecar, especially of its winds and photoionization properties.
\end{abstract}




\section{Introduction}

The massive binary system, Eta Carinae (\ecar)\footnote{Based on observations made with the NASA/ESA Hubble Space Telescope. Support for Program numbers 12013, 12750, 12508, 13054 and 13395 was provided through  grants from the Space Telescope Science Institute, which is operated by the Association of Universities for Research in Astronomy, Incorporated, under NASA contract NAS5-26555.}, has fascinated astronomers since it brightened in the 1840s to rival Sirius, faded below threshold of the unaided eye, brightened again in the 1890s, and again faded \citep{Davidson97}. In recent decades, \ecar\ has gradually brightened  to  naked-eye visibility\footnote{See  photometry at http://etacar.fcaglp.unlp.edu.ar/} \citep{lajus09}. \ecar\ is sufficiently nearby \citep[$D=2.3 \pm 0.1$ kpc,][]{Walborn12} that associated structures can be studied with moderate spatial resolution throughout the electromagnetic spectrum. 
\cite{Damineli96}  discovered a period of 5.5 years based upon the periodic modulation of the He~I 10830\AA\footnote {All wavelengths in this article are referenced to vacuum.} plus other He~I, [Ne~III], [Ar~III] and [Fe~III] lines.  He suggested a massive primary, \ecar~A, with a hotter, less massive, secondary star, \ecar~B, in a highly eccentric orbit, could account for the observed high- and low-ionization states. While \cite{damineli97} estimated the binary orbit eccentricity to be 0.63, \cite{Davidson97a} suggested that eccentricity had to be at least 0.8 while \cite{corcoran01} derived an even higher e$=$ 0.9.

\cite{Pittard02} modeled the X-ray spectrum obtained with the \chandra\ X-ray grating and determined that the wind of \ecar~B required  terminal velocity, \vinf$_{,B}=$ 3000 \kms, with a mass loss rate, $\mdot_B= 10^{-5}\msunyr$ while the wind of the much more massive star, \ecar~A, was in the range of \vinf$_{,A}=$ 500 to 700 \kms\ and $\mdot_A = 2.4 \times\ 10^{-4}\msunyr$. \cite{Okazaki08} were able to reproduce the {\it Rossi X-ray Telescope Explorer} (\rxte) X-ray light curve \citep{corcoran01, corcoran05} using an orbital eccentricity of 0.9. \cite{parkin09} varied the eccentricity and wind parameters and concluded that an eccentricity close to 0.9, but likely less than 0.95, fitted the \rxte\ X-ray light curve published by \cite{corcoran05}. While the best estimates of the secondary wind have not changed, improved models for \ecar~A's spectrum suggest  \vinf$_{,A}= $ 420 \kms\ and $\mdot_A=$  8.5$\times$10$^{-4}\msunyr$\ \citep{Groh12}.

 Previous speckle interferometry of \ecar\ by \cite{weigelt86}  and speckle masking by \cite{Hofmann88} resolved four point-like sources within 0\farcs3 of each other (called Weigelt A, B, C and D).  \cite{Weigelt95} used {\it Hubble Space Telescope/Faint Object Camera} \hst/\foc\ observations to confirm the existence of these four objects in the UV as well as the visible regions. \cite{Falcke96} obtained speckle polarimetric imagery at H$\alpha$\ of \ecar\ that were consistent with an elongated, disk-like structure extending from northeast to southwest across \ecar.  \cite{davidson1995}, using the \hst{\it/ Faint Object Spectrograph} (\hst/\fos), determined that Weigelt A is a stellar source, but Weigelt B, C and D proved to be extraordinarily bright emission-line sources located within 0\farcs3  (projected 690 AU) of \ecar\footnote {Teodoro et al. (in prep) demonstrate that the Weigelt objects are UV-illuminated surfaces of large, massive, slowly-moving clumps of infrared-bright material noticed by \cite{chesneau05}.}. 
 
Discovery of the 5.5-year period by \cite{Damineli96} coincided with installation of the {\it Space Telescope Imaging Spectrograph (\stis)} in  \hst, which enabled near diffraction-limited spatial resolution (0\farcs1 at 5000\AA) spectroscopic studies  at a selection of optimal spectral resolving powers ranging from a few hundred to a few hundred thousand \citep{Kimble98, woodgate98}. 
\cite{zethson01} and  \cite{ Zethson12} used  \hst/\stis\ spatially resolved, time variable spectra of Weigelt B and D to identify a few thousand nebular lines primarily of ionized iron-peak elements. During the 5.2-year, high-ionization state, emission lines from many species, most notably iron-peak elements were identified originating from singly to triply ionized states. Strong lines  of [Ne III] demonstrated that far ultraviolet (FUV) emission, with energies exceeding 40 eV, excited the Weigelt clumps. However, during the months-long low-ionization state, the FUV radiation drops and the spectrum is dominated by singly ionized species.   

Observations in the ultraviolet were first accomplished with the {\it International Ultraviolet Explorer (\iue)}
\citep{Heap78} and extended by \hst/\stis\ \citep{Nielsen05A, gull06} and the {\it Far Ultraviolet Spectroscopic Explorer (\fuse)} \citep{iping05}. Both the \hst\ and \fuse\ observations confirmed that the low-ionization state was indeed brought about by a drop in the ultraviolet flux most noticeably below 1500\AA. Indeed, the nearly one thousand ultraviolet narrow absorption lines of molecular hydrogen seen at $-$512 \kms\ across the high-ionization state, due to the expanding Homunculus, disappeared during the 2003.5 low-ionization state and reappeared as the high-ionization state again resumed \citep{Nielsen05A}. \cite{hillier01, hillier06} modeled the \hst/\stis\ and \fuse\ observations showing that while the spectrum of the primary star dominates from the FUV to longer wavelengths, the intrinsic obscuration in the \ecar\ system and the greatly extended primary wind prevented direct detection of the secondary companion. Hence while the periodic modulation of the X-rays, the visible-wavelength broad emission lines and the FUV demonstrate the presence of a hot secondary companion, direct detection is yet to be accomplished.

A very peculiar property of \ecar's spectrum is the presence of both narrow ($\approx$40 \kms FWHM) and broad components ($\approx$900 \kms) for many permitted and forbidden emission lines as measured by seeing-limited (1-2\arcsec) ground-based spectroscopy.  Spectra recorded by   \hst/\stis\ with the $52\arcsec\times\ 0\farcs1$\ aperture showed a broad line profile at the stellar position and a narrow line profile originating from Weigelt objects C and D \citep{gull09}. However weak narrow, velocity-shifted components could be seen extending beyond \ecar\ at various position angles (PAs). The strengths of these components changed between \hst/\stis\ observations recorded at different orbital phases of \ecar\ across the interval from 1998.0 to 2004.3. 

Also peculiar is the anomalous reddening towards \ecar\ \citep{hillier01}. In addition to normal interstellar reddening in the direction of the Carinae complex, a relatively grey extinction is thought to originate within the Homunculus and more likely close to the star. Curiously this extinction does not appear in the spectra of the Weigelt clumps \citep{gull09, mehner10}, indicating that the grey extinction possibly is circumstellar.

Extensive reviews of the many \hst/\stis\ longslit spectra of \ecar, recorded from 1998 to 2004, led to the realization that extended wind structures could be traced along multiple PAs, and that the contributions of low- and high-ionization forbidden lines (e.g. [Ne~III], [Fe~III], [Ar~III], [Fe~II] and [Ni~II]) changed with binary phase \citep{gull09}. However, a controller board in the \stis\ failed in 2004, so no further spectroscopic studies could be done with the \hst\ until repairs were accomplished during the Servicing Mission 4 in May 2009. High spatial resolution, combined with moderate spectral resolution, using ground-based facilities was, and continues to be, not possible as adaptive optics correction, coupled with moderate spectral resolution (R$\approx$10,000), does not yet extend into the visible spectral region.

The \hst/\stis\ longslit spectra and  \hst/{\it Wide Field Planetary Camera 2}\ (\wfpc2) direct imagery revealed that the Homunculus, ejecta from the 1840s Great Eruption of \ecar, is a bipolar structure tilted about 45 degrees into the sky plane \citep{davidson01}.  However, it was not obvious that the binary orbital plane was, or was not, aligned with the Homunculus axis of symmetry. Three-dimensional (3-D) modeling of the X-ray light curve led to evidence that the axis of the orbital plane was tilted at about 45\degr\ into the sky plane, but a degeneracy existed as to which direction the tilt axis was oriented within the sky plane \citep{Okazaki08}. Improved 3-D modeling \citep{Madura12} combined with the \hst/\stis\ longslit observations showed that the normal to the binary orbital plane was closely aligned to the Homunculus major axis of symmetry, thereby suggesting the possibility that formation of the Homunculus was influenced by the binary system \citep{gull09}. The 3-D Homunculus model of H$_2$ near infrared emission strongly supports this idea \citep{Steffen14} .

Many changes with binary phase are apparent in the spectra. The most prominent is the disappearance of the high-ionization emission ($>$13.6 eV) across the periastron and with the appearance of low-ionization emission ($<$13.6 eV). These resolved emission structures are the result of not just the current wind-wind interactions, but also wind-wind interactions from the most recent cycles. Their visibility is the result of absence or presence of FUV radiation impinging upon structures with relatively high electron densities ($\approx 10^7\ cm^{-3}$). These structures are the compressed, expanding-wind structures that persist for more than one 5.5-year cycle. Hence we call them `fossil wind' structures. They  provide a historical record of photoionization, expansion, and density over the several most recent cycles \citep{Teodoro13}.

Many  observations accomplished over the past few decades show that \ecar\ is a unique, massive binary system: \begin{enumerate}
\item The interacting winds are so massive that the forbidden lines from Fe$^{++}$ and Fe$^+$  trace portions of  dense fossil shells  \citep{gull09}. 
\item The orbit is sufficiently eccentric that the hot secondary spends considerable time near its apastron position, effectively clearing a large cavity out of the primary wind. 
\item The high-ionization emission is mostly blue-shifted while the low-ionization emission is mostly red-shifted, which supports 3-D models \citep{gull09, Madura12} with apastron occurring with \ecar~B positioned on the near side of \ecar~A. Periastron occurs with \ecar~B on the far side of \ecar~A \citep{Gull11}.\end{enumerate} These fossil shells provide powerful clues to the physics of the wind-wind interactions, the orientation of the binary orbit and provide a measure of changes in the properties of the winds over several cycles. 

The \hst/\stis\ spatially-resolved spectra centered on \ecar\ recorded prior to 2004 provided individual slices of the 3-D fossil wind structures at a number of orbital phases but were insufficient to describe the entire structure and the dynamical changes across the 5.54-year cycle \citep{gull09}. Mapping was needed since single spectra at arbitrary roll angles at different orbital phases could not distinguish spatial- from phase-variability. We determined that mapping the structures revealed by selected forbidden emission lines with the \hst/\stis\ across the central 2\arcsec\ $ \times\ $ 2\arcsec\ at selected phases could provide much insight on the density and ionization conditions of these expanding shells partially photo-ionized by the FUV radiation from \ecar~B.

%

The spatial mappings of the central region surrounding \ecar\ have already  proven to be very useful for related studies:
\begin{enumerate}
\item Observations recorded from 2009 through 2013 have been used to measure expansion of structures seen in [Fe~II] and [Ni~II]
\citep{Teodoro13}.  These low-ionization structures are red-shifted, moving at 470 \kms\ and are associated with the primary wind. Gaps between the shells are due to the passage of the hot secondary across the periastron event that occurs every 5.54 years. While the primary wind velocity was measured to be 420 \kms\ \citep{Groh12}, these shells expand at a slightly higher velocity possibly due to acceleration by the much faster secondary wind.

\item The 4706\AA\ grating setting included the He~II 4686\AA\ line that dramatically strengthens as the stars approach periastron \citep{steiner04}. \cite{Teodoro16} resolved the apparent discrepancy between published ground-based and \hst/\stis\ measures \citep{davidson15}. The He~II emission comes from within  0\farcs13 centered on \ecar, but seeing-limited ground-based spectroscopy includes 1$-$2\arcsec\ of nebular structure, leading to increased continuum and a smaller He~II 4686\AA\ equivalent width measured from the ground.

\item Studies of He~I 4714, 5876 and 7065\AA\ line profile variations with ground-based, high-dispersion (R$=\lambda/\delta\lambda=90,000$) spectra were shown to be consistent with spectral profiles integrated from \hst/\stis\ mappings in the same time frame \citep{Richardson16}. While most of the He~I line profiles come from the central 0\farcs13 core, the line profiles as observed from the ground are considerably modified by absorptions and scattered light from the fossil wind structures plus slowly moving debris (the Weigelt blobs) from the two historical ejections.
\end{enumerate}

These observations show quite complex structures that change continuously across the 5.54-year binary cycle. Patterns emerge that provide clues to how the interacting winds are shaping their environment, including interactions with slowly moving clumps. Explanation demands 3-D modeling efforts to replicate the fossil wind structures. Models fitted to the observations will improve constraints on the stellar winds and the binary orientation parameters, possibly even new insights on the physics of the outbursts. 

Preliminary hydrodynamic models with limited radiative transfer  have already been accomplished by \cite{madura13} and \cite{Clementel14, Clementel15a, Clementel15b}. Additional observations at higher angular resolution in the near infrared  and radio will provide considerable new information. Specifically, observations with the {\it Very Large Telescope Interferometer (VLTI)} with angular resolutions of a few milliarcseconds are already revealing information on the current interacting wind properties (Weigelt et al. submitted).

We describe the observations  in Section \ref{obs}. Changes in structure across the series of mappings are described in Section \ref{pha}.   Discussion of these changes and implications on the properties of the two massive components are presented in Section \ref{results} with conclusions in Section \ref{conclude}.
Details of the data reduction are presented in Appendix \ref{reduce}.  Two very large mosaics of [Fe~III] and [Fe~II] showing the iso-velocity slices for mappings are presented in Appendix \ref{global}.  while  a discussion of the changing fluxes is given in Appendix \ref{flx}.

 \section{Observations} \label{obs}

The first  mapping observations for this study were done with the \hst/\stis\ in June 2009 across the $2\arcsec \times 2\arcsec$\ region centered upon \ecar\ with the $52\arcsec \times\ 0\farcs1$\  aperture at $0\farcs1$  spacing (230 AU at 2.3 kpc distance)  for several grating settings including the G430M grating centered at 4706\AA\ \citep{Noll09}.
 Each of the grating settings had been selected to  map specific forbidden lines that would reveal a range of ionization states. We had reviewed the line list, provided  by \cite{zethson01}, searching for strong forbidden lines that were minimally contaminated by other emission lines or absorptions from strong wind lines such as He~I 4714\AA, which is on the red side of [Fe~III] 4701\AA. Two forbidden lines of interest, [Fe~III] 4659.35\AA\ and [Fe~II] 4815.88\AA, both of which are relatively isolated from wind and nebular lines, proved to be excellent candidates as they originate from the same atomic species and provide ionization information above (16.2 eV) and below (7.9 eV) the ionization potential of hydrogen (13.6 eV). 
 
Other grating settings in the initial visits provided mappings of [S~III] 9071, 9533\AA\ and 6313\AA, [N~II] 5756\AA, [Ar~III] 7137\AA\ and [Fe~II] 7157, 7173\AA\ \citep{Noll09, Corcoran09}. The [S~III] mapping proved to be of less utility  due to the \stis\ CCD performance and the broadened \hst\ point spread function in the near red. Given  the very limited \hst\ observing time, we streamlined the full cycle program  to obtain maps  of [Fe~III] 4659\AA\ and [Fe~II] 4815\AA\ at selected phases, but we added maps of [N~II] 5756\AA\ and [Ar~III] 7137\AA\ at phases when considerable changes in ionization were predicted by the 3-D models \citep{madura10}. This article focuses on the changes of [Fe~III] 4659\AA\ and [Fe~II] 4814\AA,  herein labeled [Fe~III] and [Fe~II], across a full orbital cycle. Mappings of [N~II], [Ar~III] and He~I will be presented in other papers.

\begin{table}
\caption{\hst/STIS Data Used in This Analysis}
\label{t1}
\begin{tabular}{lllll} \hline
\hst\ Obs. &Obs. Date &  Orbital$^{a}$  & Slit PA & Mapped \\
Prog& &   Phase& Degrees&Region$^{b,c,d}$\\ 
\hline
11506$^{e}$ & Jun 29, 2009 & 12.08408 & $+$79.5 & 3\farcs2 $\times$ 2\farcs0\\
12013$^{e}$ & Dec 6, 2009&12.16318 & $-$121.0 &3\farcs2 $\times$ 1\farcs4\\
12013$^{e}$ &  Oct 10, 2010& 12.31545 & $-$166.7 & 6\farcs4 $\times$ 1\farcs1\\
12508 & Nov 20, 2011 & 12.51617 & $-$138.7 & 6\farcs4 $\times$ 2\farcs0\\
12750 & Oct 18, 2012 & 12.68080 & $-$174.8 & 6\farcs4 $\times$ 2\farcs0\\
13054 & Sep 3 2013 & 12.83901 & $+$136.7 & 6\farcs4 $\times$ 1\farcs8\\
13395  & Feb 17, 2014 & 12.92157 & $+$169.2& 6\farcs4 $\times$ 1\farcs9\\
13395 &Jun 9, 2014 &12.97694&$+$61.4&6\farcs4 $\times$ 2\farcs0\\
13054 &Aug 2, 2014 &13.00364&$+$107.7&6\farcs4 $\times$ 1\farcs8\\
13054 &Sep 28, 2014 &13.03182&$+$162.1&6\farcs4 $\times$ 1\farcs8\\
13395$^{f}$ & Dec 16, 2014 & 13.07088 & $-$113.2 & 6\farcs4 $\times$ 1\farcs8\\
13395& Jan 16, 2015 & 13.08620 & $-$85.7 & 6\farcs4 $\times$ 2\farcs0\\
13395& Mar 13, 2015 & 13.11389 & $-$31.8 & 6\farcs4 $\times$ 1\farcs8\\
 \hline
\end{tabular}\\
$^{a}$ Phase, $\phi$, refers to JD=JD$_o$+2022.7($\phi$-11), where JD$_o$=2,452,819.2, corresponding  to when the  narrow component of HeI $\lambda6678$  disappears \citep{groh04, damineli08_period}. Note that  periastron is thought to occur about ten days later in the 5.54-year cycle based upon modeling of  observations of He~II 4686\AA\ \citep{Teodoro16}.\\
$^{b}$ Map sampling during the first two visits were 0\farcs1 intervals perpendicular to the STIS 52\arcsec $\times$ 0\farcs1 aperture. All other mappings were accomplished at sampling with 0\farcs05 intervals.\\
 $^{c}$ Extent of the mapped regions was limited by the 96 minute \hst\  orbit since all observations were accomplished in continuous viewing orbital opportunities.\\
$^{c}$ All observations discussed in this article were performed with the {\it STIS} grating, G430M, centered at 4706\AA.\\ 
$^{e}$ Observations during the first three visits were done with non-CVZ orbits and hence were constrained to 20 spectra per orbit in order to fit within the \hst\ orbital observing window.\\
$^{f}$ Observations for the December 2014 were broken into two normal \hst\ orbits due to a guide star failure during the planned single CVZ orbit scheduled for late November 2014.\\
\end{table}

The first \stis\ mapping occurred five months after the 2009.0 periastron event and confirmed that \ecar\ and the excitation of its fossil winds were returning to the high-ionization state as demonstrated by seeing-limited, ground-based observations done in the same time period \citep{Richardson15}. Additional mappings (see Table \ref{t1}) were accomplished through guest investigator programs \citep{Corcoran11,  Gull11b, Gull12, Gull13}. Variations in sampling frequency (0\farcs1 versus 0\farcs05) and mapped area (1\arcsec\ versus 2\arcsec) demonstrated that optimal sampling is one-half the width of the \stis\ aperture,  0\farcs05, and that the size of the fossil wind structure required a minimum mapped area of 1\farcs8 in diameter. 

When possible, as \ecar\ lies in the direction of the \hst\ orbital pole, we requested  continuously viewing zone (CVZ) orbits. This allowed us to double the number of acquired spectra per orbit. A full map at each grating setting was accomplished within an individual orbit without visibility interruption by Earth passage.  Most visits were accomplished with CVZ orbits. Exceptions were  the first three visits done in June 2009, December 2009 and October 2010, plus a replacement visit in December 2014 due to loss of a guide star during the planned November 2014 visit. The November/December 2014 visit was very critical  as we needed information on the early recovery structures after the 2014.6 periastron event. The CCD readout was limited to 128 rows (6\farcs3) to minimize readout overheads.

Orientation of the 52\arcsec $\times$\ 0\farcs1 aperture was limited by constraints imposed by  \hst\ solar panel orientation, so the PA could not be specified. All mappings were resampled to a common right ascension/declination grid centered on the position of \ecar. 

Earlier ground-based \citep{Richardson15},  \hst/\stis\ observations \citep{gull09} and the 3-D models of the interacting winds \citep{madura10, Gull11, Madura12, madura13} demonstrated that structural changes were slow across the long, high-ionization state (apastron) when the binary members are furthest apart and moving slowly.  \hst/\stis\ visits  were scheduled at one year intervals  from 2010 to 2013. 

Visits were much more frequent before, during and after the periastron event. Across the short, low-ionization state (periastron), the stars approach to within a few AU, moving at velocities relative to each other that approach \vinf$_{,A}$ \citep{Nielsen07}.  From the 3-D models, we selected critical phases where structures were expected to change in ionization, which determined critical times between February 2014 and March 2015 when observations were needed (see Table \ref {t1}). 

The timing of these observations was selected based upon the known changes from high-ionization to low-ionization state. \cite{damineli08_period} and \cite{damineli08_multi} reference $\phi=$ 0 to the disappearance of the narrow, emission component for He~I 6678\AA. The present observations sample the high-ionization state at multiple intervals, $\phi=$ 12.163 through 12.977, 13.071 through 13.114, but only twice during the low-ionization state, $\phi=$  13.004 and 12.032. While the drop to low-ionization state is well identified, observations of the recovery from low- to high-ionization are very limited during past periastron passages.  The most recent recovery was not covered from the ground due to limited night-time visibility. Hopefully substantial coverage will be possible during the recovery phase of the 2020.1 periastron passage as it is well-placed in southern winter.

Data reduction is described in Appendix \ref{reduce}.

\section{Structural changes with orbital phase}\label{pha}
Here we describe changes in structures revealed in the light of [Fe~III] and [Fe~II] from $\phi=$ 12.084 to 13.114, a span of 5.7 years extending  from early recovery of the low-ionization state  across a high-ionization state to nearly full recovery after the next low-ionization state associated August 2014 periastron passage. These changes are predominantly caused by two  effects: 1) changes in structures due to photo-ionization and 2) radial expansion of the wind structures with time. 

For the convenience of the reader, \ecar\ is at a distance of 2300 pc \citep{Walborn12}. An angular dimension of 0\farcs1 translates to 230 AU. Over the 5.54-year orbital period, 500 \kms\ translates to an spatial motion of 516 AU or 0\farcs2 on the sky plane.

We first describe the major changes from apastron to periastron in Section \ref{minmax}.  A global view with highlights is presented in Section \ref{global} along with full mosaics of the [Fe~III] (Fig. \ref{FeIIImosaic} and the [Fe~II] (Fig. \ref{FeIImosaic}) iso-velocity images.
A brief description of long term changes since 1985 is described in Section \ref{long}. Changing structures are described in Section \ref{changes} with focus on 
the slowly expanding structures, including the Weigelt objects (Section \ref{slow}), the rapidly approaching structures  (Section \ref{blue}) and the rapidly receding structures (Section \ref{red}). A novel means of measuring structural changes is described in Section \ref{expandlight}. The technique is applied to [Fe~III] in Section \ref{light} and to [Fe~II] in Section \ref{expand}.  
\subsection{The extreme changes from apastron to periastron}\label{minmax}
\begin{figure*}
\includegraphics[width=6.5in,angle=0]{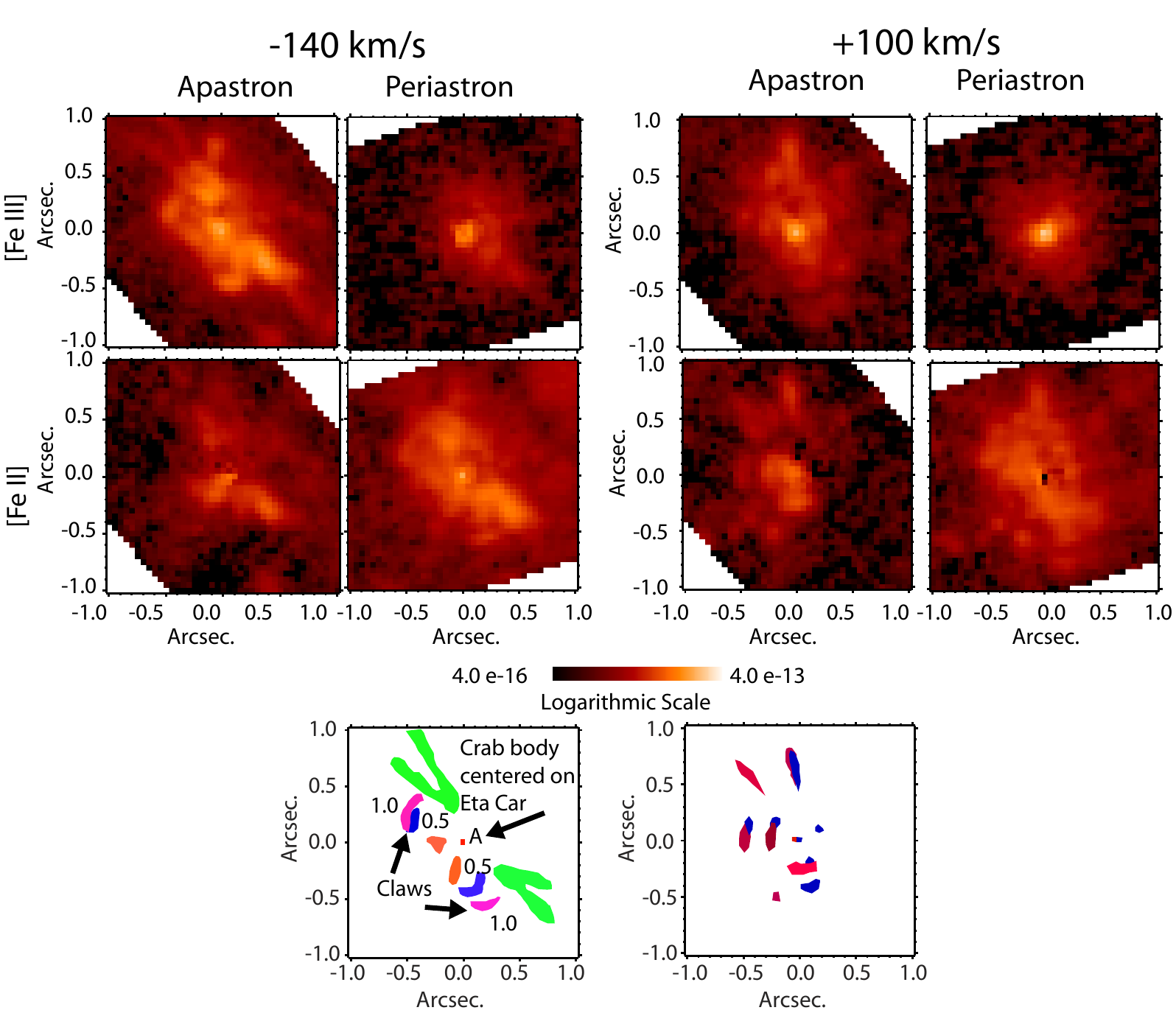}
\caption{{\bf Changes between apastron ($\phi=$ 12.516) and periastron ($\phi=$ 13.004). Top Left:} Comparison of  $-$140 \kms\ images. Structures seen in [Fe~III] at apastron (upper left) appear in [Fe~II] at periastron (lower right), but having expanded radially from \ecar\ over the 2.77 year interval. {\bf Top Right:} Comparison of $+$100 \kms\ images between apastron and periastron. Again structures seen in [Fe~III] at apastron are seen in [Fe~II] at periastron, having expanded over the time interval. {\bf Bottom Left:} Sketch depicting selected  $-$140 \kms\ filaments seen in [Fe~III] at apastron and in [Fe~II] at periastron.  Arcuate structures (blue) seen in [Fe~III] expanded outwardly and appear in [Fe~II] (pink) at periastron. New arcuate structures (orange) are seen in [Fe~III] at apastron and in [Fe~II] at periastron. Bifurcated structures (green) seen in [Fe~III] at apastron are not obvious in [Fe~II] at periastron.  {\bf Bottom Right:} Sketch of $+$100 \kms\ structures seen in [Fe~III] (blue) and in [Fe~II] (red) at apastron. Most structures show interior [Fe~III] and exterior [Fe~II] components, consistent with photoionization from the central source. Expansion in [Fe~III] across the high-ionization state is consistent with 450 to  470 \kms, close to the \vinf$_{,A}=$  420 \kms\ derived by \protect\cite{Groh12} and the 470 \kms\ expansion velocity of [Fe~II] structures as measured by \protect\cite{Teodoro13}.  However, shifts in illumination due to the motion of the secondary (FUV source) star, intervening structures that first absorb 
FUV radiation, then break open, and expansion of the shells lead to rotational shifts of the [Fe~III] features: i.e. different portions of the shells  emit [Fe~III] as the orbit progresses. By contrast the arcuate features  emitting [Fe~II] appear to move outward radially (note: all images in this figure have continuum subtraction).
 \label{yinyang}}
\end{figure*}

The most noticeable contrast in emission structures revealed by [Fe~III] and [Fe~II] is between apastron, when the two stars are separated by  the orbital major axis, $\approx$30 AU, and periastron, when the separation drops to $\approx$1.5 AU (Fig. \ref{yinyang}). The 3-D hydrodynamic models \citep{Okazaki08, Pittard02, parkin09, parkin11, Madura12, madura13, Clementel15a, Clementel15b} demonstrate that near apastron, the cavity carved by \ecar~B out of the primary wind is open and nearly parabolic, allowing FUV, originating from \ecar~B, to escape the current wind-forming cavity\footnote {While the predominant source of FUV radiation is thought to be \ecar~B,  the colliding winds, the source of X-rays which reach a shallow minimum across periastron, may also contribute. However we would expect the FUV radiation from the wind-wind collision zone to be least near apastron when the two stars are most distant and to increase with approaching periastron. A complication comes up with evidence that as periastron approaches, the winds reach a stalling point \citep{Clementel15b}. Additional modeling is necessary to determine the contribution by the colliding winds.}.

 As the system approaches periastron, the distance between the two stars decreases, and, since the distance of the wind-wind shock`wall' from \ecar~B scales with  the stellar separation, the shock boundary moves closer to \ecar~B.  By periastron, \ecar~B's orbital velocity becomes comparable to the primary wind terminal velocity. The cavity becomes greatly distorted with the leading edge of the bowshock penetrating deeply into the primary wind and with the trailing edge  far behind the secondary wind. Rapidly, FUV radiation from \ecar~B is blocked by the slow-moving primary wind. Hydrogen in the primary wind absorbs the FUV, converting it to radiation with energies  less than 13.6 eV and an optically thick Lyman $\alpha$ radiation field (10.4 eV). The high-ionization state is no longer supported external to the very shrunken wind-wind cavity.  The mid-ultraviolet radiation, $\approx$ 10 eV to 5 eV (MUV), which comes from both the primary and the converted FUV from the secondary, escapes the wind-wind cavity. Throughout the fossil wind structures, iron formerly as Fe$^{++}$ drops to Fe$^+$ and many [Fe~III] structures reappear as [Fe~II] structures.

These changes are driven by UV radiation, absorption of the UV radiation, expanding structures and densities of the excited gas. Iron is ionized to Fe$^+$ by MUV exceeding 7.9 eV and further ionized to Fe$^{++}$ by FUV above 16.2 eV. Thermal excitation populates these ions to the upper state of the transitions in question, which occur at 2.7  eV with critical densities for the transitions of interest: [Fe~III] 4659.35\AA: n$_e=$  10$^7$ cm$^{-3}$ and [Fe~II] 4815.88\AA: n$_e=$ 2$\times$10$^6$ cm$^{-3}$.

The extremes of the high-  and low-ionization states are compared for $-$140 \kms\  and  $+$100 \kms\ velocities in Fig. \ref{yinyang}.  The [Fe~III] structures, visible at apastron,  disappear  at periastron only to reappear as expanded, more diffuse, [Fe~II] structures.  The  periastron [Fe~II] structures are enlarged relative to the apastron [Fe~III] structures due to radial expansion during the 2.7-year time interval between $\phi=$ 12.516 and 13.004.

This effect is most noticeable in the $-$140 \kms\ images in Fig. \ref{yinyang}. As drawn in the cartoon (bottom left) two  [Fe~III] arcs, visible at apastron (labeled 0.5), shift outward and change to   [Fe~II] arcs, visible at periastron (labeled 1.0).
Two pairs of very extended [Fe~III] structures (green in the left cartoon) are not seen in [Fe~II] other than structures closest to \ecar. In general, periastron-visible [Fe~II] structures spatially project to the outside of the apastron-visible [Fe~III] structures.

The contrast is less for $+$100 \kms\ structures displayed in Fig. \ref{yinyang}. Less [Fe~III] emission is apparent at apastron as these structures are on the far side of \ecar\ where little FUV radiation can penetrate. Close to \ecar\ are several arcuate [Fe~III] structures that form a nearly complete 0\farcs6 diameter ring, plus a faint streak that projects directly north (blue in the cartoon in Fig. \ref{yinyang}, right).  The [Fe~II] structures at apastron  clearly define two sets of arcs, plus a pair of streaks, one associable with the [Fe~III] streak to the north and a second to the northeast. The [Fe~II] emission at periastron reveals many additional arcs/clumps at distances out to nearly 1\farcs0. These additional arcs are likely structures that, at apastron, are ionized by the FUV radiation but have densities lower than n$_e=$ 10$^7$ cm$^{-3}$, comparable to n$_e=$ 2$\times$10$^6$ cm$^{-3}$, the critical density for the [Fe~II] 4815.88\AA\ emission. Hence they are not seen in [Fe~III] at apastron, but in [Fe~II] at periastron.

\subsection {Long term changes since 1985} \label{long}

\begin{figure*}
\includegraphics[width=6.5 in, angle=0]{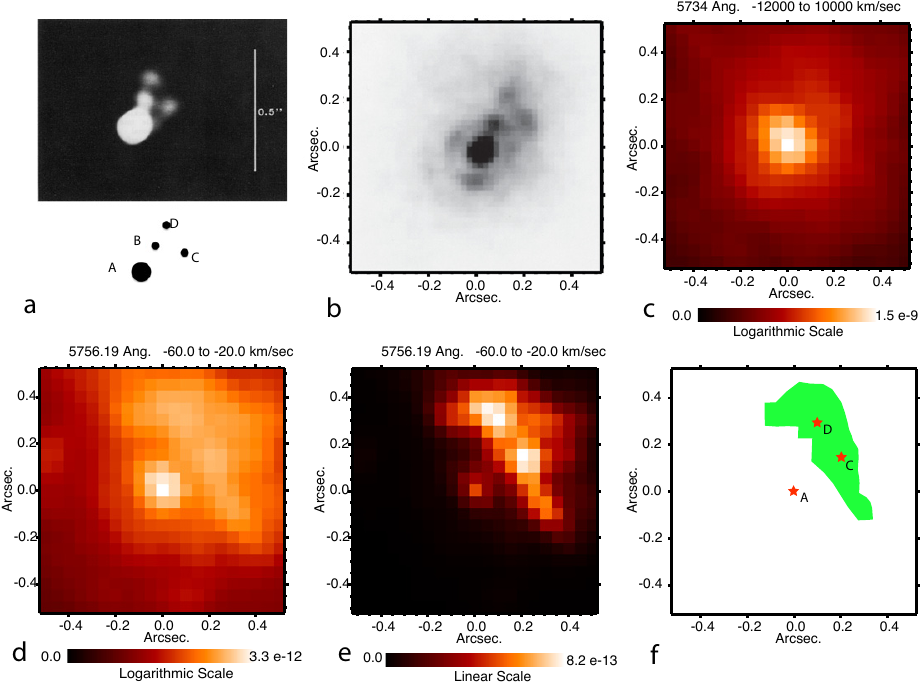}

\caption{{\bf Comparison of the current structures and how they relate to the original Weigelt objects.}  {\bf a: Speckle image recorded in 1985 as published by \protect \cite{Hofmann88}.} Spectral response was from 8300 to 8800\AA. Note the sketch at bottom showing four point-like objects. A is \ecar. B, C and D are very bright emission clumps known as the Weigelt blobs. {\bf b: \hst/{\it FOC} image as published by 
\protect \cite{Weigelt95}}.
 The spectral bandpass was defined by the {\it FOC}\ F550M filter and S20 photocathode response. {\bf c: \hst/\stis\ image integrated from 5500 to 5920\AA,} the full range of the   G750M grating recorded on November 2011 (apastron, $\phi=$ 12.516). Little or no evidence of Weigelt objects B, C or D is obvious in this broadband image. {\bf d:  \hst/\stis\ image from the same mapping with the bandwidth constrained to 40 \kms\ centered at $-$40 \kms\ for [N~II] 5756.19\AA.} The  bright central component is \ecar. The extended, hook-shaped structure to the northwest is in the vicinity of Weigelt C and D. Continuum has {\bf NOT} been subtracted from either frames c or d. {\bf e: The continuum-subtracted \hst/\stis\ image with 40 \kms\ bandwidth centered at $-$40 \kms\ for [N~II] 5756.19\AA.} Faint [N~II] emission is centered at coordinates 0,0 which is the position of \ecar. Weigelt B is not visible at any of the bright emission lines since these mappings began in 2009. {\bf f: Sketch depicting the positions of \ecar\ (A), Weigelt C and D, plus the hook-shaped nebular structure that extends around C and D.}   
As noted by Teodoro et al. (in prep, 2016) the peak nebular emission positions in the current epoch do not correspond to previously measured positions of B, C and D due to proper motion, changes in photoionization, illumination and possible ablation. The change from the three clumps of emission, detected in 1985 and 1995, to the more diffuse structures by 2011 suggests that all three Weigelt objects are slowly evolving and dispersing.
Note that these images are dimensioned 1\arcsec $\times$ 1\arcsec. All other images in this article are 2\arcsec $\times$ 2\arcsec. Images presented in c.-e. have been logarithmically scaled as shown in each color wedge to maximize displayed range. Images a and b are used with permission of the authors.\label{field}}
\end{figure*}

In the time since \cite{weigelt86} and \cite{Hofmann88}  resolved \ecar\ plus three point-like sources within 0\farcs3 via ground-based speckle interferometry (Fig. \ref{field}.a), the apparent brightness of the central sources has increased by nearly a factor of three at visible wavelengths, part of a  trend since 1940  \citep{davidson99a, lajus09}. A decade later \cite{Weigelt95} via the \hst/{\it FOC} confirmed these stellar-like sources persisted (Fig. \ref{field}.b). \cite{davidson95} obtained observations with \hst/{\it FOS} that demonstrated the spectrum of Weigelt A is stellar-like  with a hot, dense wind, while objects B, C and D are ejecta with many forbidden lines.  Proper motion studies by \cite{Weigelt95} and \cite{  Smith04} suggested these objects were ejected between 1880 and 1930 while \cite{Dorland04} found ejection between 1900 and 1950. More recently \cite{weigelt12} found a proper motion of C and D of 2 mas  yr$^{-1}$  consistent with  ejection between 1860 and 1900. 

Long aperture spectroscopy with \hst/\stis\ from 1998 to 2004 showed that emission line fluxes of Weigelt D did not increase proportionately with the  flux of \ecar\ \citep{hillier92, hillier01, gull09, mehner10}. The apparent brightening appears to be due to changes in obscuration between \ecar\ and the observer, but not in the direction from Weigelt D to the observer. \cite{Weigelt95} suggested an elongated obscuring structure to be a circumstellar disk in front of \ecar\ at PA 50$\degr$ which is orthogonal to the symmetry axis of the Homunculus.
To study this occulting structure, \cite{Falcke96} carried out speckle-masking, imaging polarimetry in H$\alpha$. They obtained images (see their Fig. 2b) that show a northeast to southwest bar in polarized light  about 0\farcs4 long oriented at PA$\sim$45$\degr$. This observed northeast to southwest bar was interpreted to  be an equatorial disk. The polarized bar has approximately the same length and PA as that of the extended emission structures discovered by \cite{Gull11}.

Comparisons are presented in Fig. \ref{field} of images recorded by speckle masking in the 1980s (Fig. \ref{field}.a), by \hst/{\it FOC}\ in the mid-1990s (Fig. \ref{field}.b) and by \hst/\stis\ mappings in 2011 (Fig. \ref{field}.c-e). Integration of the \stis-mapped data across the complete spectrum recorded with the G750M grating centered at 5734\AA\ (Fig. \ref{field}.c) do not readily show the Weigelt objects. The dominant structure appears to be a diffuse structure in the diffraction ring region, plus a streak along the aperture when placed directly on \ecar. Constraining the velocity of the image to $-$40 \kms\ with width of 40 \kms\ (comparable to the instrumental resolving power, R$=\lambda/\delta\lambda=$ 8000), centered at [N~II] 5756.19\AA. Fig. \ref{field}.d reveals a hook-shaped structure offset to the north and west from \ecar. When continuum is subtracted, the hook structure becomes quite apparent (Fig. \ref{field}.e). The two brightest peaks in  Fig. \ref{field}.e correspond closely to the positions of Weigelt C and D, given a proper motion of 2 mas yr$^{-1}$ as measured by \cite{weigelt12}. For the reader's benefit, a cartoon (Fig. \ref{field}.f) illustrates the positions of Weigelt C and D relative to \ecar\ at the time of the \stis\ observation. However, as noted by Teodoro et al. (in prep) the emission line shapes of Weigelt C and D change with orbital phase. Weigelt B is not detected in any mappings recorded between 2009.4 and 2014.3, likely because it has  dissipated through heating, photoionization and wind ablation effects. Furthermore, the structure of Weigelt C and D are slowly evolving and eventually will dissipate. Changes in structure due to these effects must also influence proper motion studies.

\subsection{Detailed comparisons of [Fe~II] and [Fe~III] over the 5.54-year cycle}\label{changes}
Figs. \ref{a-40} through \ref{a+220} display selected velocity slices comparing the differences and changes in [Fe~III] and [Fe~II] in greater detail. All velocity slices utilize a log scale of the flux ranging from 4$\times$10$^{-16}$ to 4$\times $10$^{-13} $\ erg cm$^{-2}$sec$^{-1}$, which covers the range in flux seen in both [Fe~III] and [Fe~II] in all velocity frames. Each figure has four columns: 

The left two columns display [Fe~II] and [Fe~III] for the seven observations spanning from early recovery of the high-ionization state ($\phi=$ 12.084) to late stages before periastron ($\phi=$ 12.922). Since the orbit is very eccentric, the hot secondary spends most of the orbit  nearly 30 AU  from the primary; its distance and orbital angle change very slowly, so the photo-ionized structures evolve more slowly. The dominant changes are expansions of the fossil winds and apparent shifts in illumination of slow-moving, irregular structures which we label the  light house effect  (Fig. \ref{claws}).

The right two columns display [Fe~II] and [Fe~III] for seven observations spanning from the late high-ionization state ($\phi=$ 12.922) across the periastron-induced low-ionization state to the early high-ionization state ($\phi=$ 13.114). Note that  the maps from $\phi=$ 12.922 are displayed at both the bottom of the high-ionization state (left two columns) and top of the low-ionization state (right two columns) to enable visual comparisons. Changes are rapid across the low-ionization state. Within the resources allocated, we attempted to sample as frequently as possible, given the CVZ opportunity once every 57 days. While noticeable changes occur between $\phi=$ 12.922 and 12.977, the [Fe~III] really drops between $\phi=$ 12.977 and 13.004. Well-defined structures disappear leaving a diffuse haze at every velocity, most noticeably at $-$40 \kms\ (Fig. \ref{a-40}) for both $\phi =$ 13.004 and 13.032. By $\phi =$  13.071, [Fe~III] structures reappear in early recovery of the high-ionization state. Maps at  $\phi=$ 13.086 and 13.114 show continued recovery to the high-ionization state that can be compared to the maps at 12.084 and 12.163, mapped at half of  the optimal sampling frequency.

\subsubsection{The slowly moving  structures} \label{slow}

\begin{figure*}
\includegraphics[height=7.8in,angle=0]{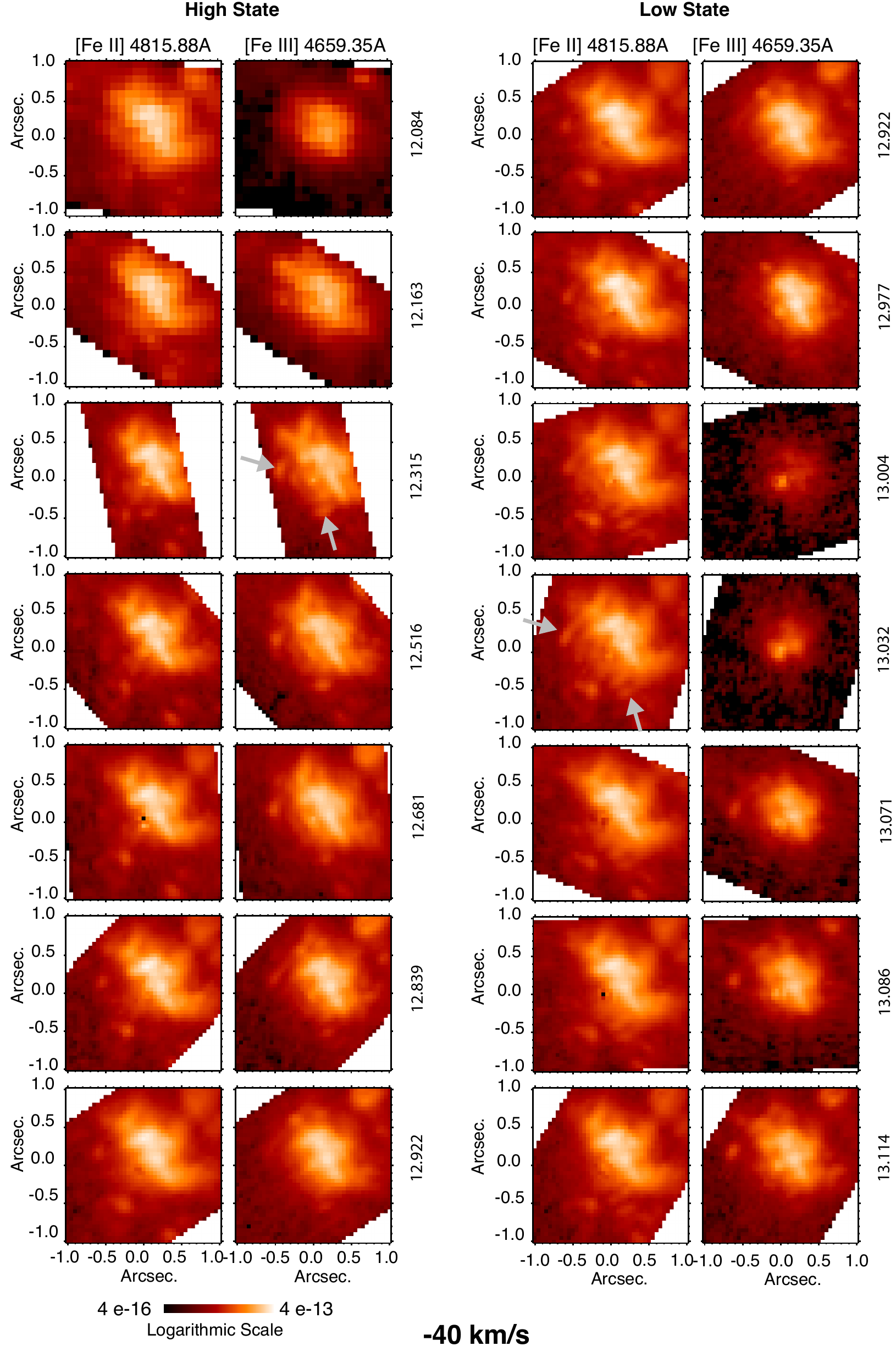}
\caption[width=3in]{{\bf Changes in spatial structure from 2009.4 to 2014.2: $-$40 \kms. Left two columns:} Iso-velocity slices of [Fe~II] and [Fe~III] for the seven mappings across the  high-ionization state from phases 12.084 to 12.922.  {\bf Right two columns:} Iso-velocity slices of [Fe~II] and [Fe~III] across the low-ionization state from $\phi=$ 12.922 to 13.114. The Weigelt objects and the hook structure (Fig. \ref{field}) change slowly in [Fe~II] but nearly disappear in [Fe~III] across the low-ionization state. Fainter, outer structures visibly shift outward from \ecar. Notable are two [Fe~III] arcuate segments that appear in [Fe~III] at $\phi=$ 12.315, one 0\farcs4 east (grey arrows, [Fe~III] at $\phi=$ 12.315), the other 0\farcs4 south of \ecar. These arcuate segments are faint at $\phi=$ 12.315,  become more prominent  by $\phi=$ 12.516 and shift radially outward across the high-ionization state.  By $\phi=$ 12.922, these  arcs have moved to 0\farcs6 distance from \ecar\  and are fading in [Fe~III]. The arcuate structures brighten in [Fe~II] (grey arrows, [Fe~II], $\phi=$ 13.032)  as the system goes through the low-ionization state ($\phi=$ 13.004 and 13.032) and are joined by two interior pairs. By recovery at $\phi=$ 13.114, the eastern arc is still visible, but the southern arc is very faint.  The $\phi=$ 12.922 mapping is displayed twice, at the bottom of the high-ionization  state columns and the top of the low-ionization state columns, as it is a transitional phase at the end of the high-ionization state and beginning of the low-ionization state. \label{a-40}}
\end{figure*}

The strongest forbidden emissions originate in Weigelt  C and D, plus associated, fainter structures in the velocity interval from $-$80 to 0 \kms\ (see Fig. \ref{a-40} for the $-$40 \kms\  iso-velocity images with 40 \kms\ width). In general, Weigelt C is slightly brighter than Weigelt D, but small shifts in relative brightness and apparent position occur across the cycle. 

The under-sampled maps at $\phi=$ 12.084 and 12.163 (Fig. \ref{a-40}) are very similar to the optimally sampled maps at $\phi=$ 13.086 and 13.114,  one orbital period later, which indicates that the low-velocity structures have not evolved much in one 5.54-year period,  that the UV sources of ionization for Fe$^{++}$ (16.2 eV) and Fe$^+$ (7.9eV) have not changed significantly nor have the densities of the current fossil structures changed much from the previous fossil structures at comparable phases one cycle earlier.

The structures of greater importance  are the much fainter outer structures that appear and disappear across the cycle:

\begin{enumerate}
\item Two arcuate segments of [Fe~III] appear by $\phi=$ 12.315: one located  0\farcs4 at PA$=$ 80\degr\ and the other located 0\farcs4 at PA$=$ 190\degr\ from \ecar (see structure indicated by arrows for $\phi =$ 12.315 in Fig. \ref{a-40}). These arcuate segments, which we name `claws' (as in crab claws), reappear in other iso-velocity slices.  In following sections, the claws are followed as they change with phase. 
\item These [Fe~III] claws expand outwardly with advancing phase but begin to fade by $\phi=$  12.922 while brightening in [Fe~II].
\item A second, fainter pair of [Fe~III]  claws appears at about the same PA, 0\farcs2 from \ecar\ but are not as obvious across the high-ionization state from $\phi=$ 12.516 onward. They reappear as [Fe~II] claws across perastron ($\phi=$  13.004 and 13.032) about 0\farcs4 from \ecar\ at PA$=$ 60 and 200\degr.
\item At $\phi=$  13.004 and 13.032 the outer claws, now visible in [Fe~II],  have expanded to $\approx$0\farcs6, are more extensive and are joined by two internal pairs at $\approx$ 0\farcs4 and 0\farcs2. The three pairs of arcuate segments indicate repeated formation of shells each cycle.
\item By $\phi=$ 13.071, a single [Fe~III] claw at 0\farcs6, PA=80\degr\ is again visible, but the claw at 0\farcs6, PA$=$ 190\degr\ has not reappeared. 
\item No additional claws are visible at $\phi=$ 13.114, similar to no claws being  visible at $\phi=$ 12.163.  

\end{enumerate}

\subsubsection{The blue-shifted structures}\label{blue}

\begin{figure*}
\includegraphics[height=7.8in,angle=0]{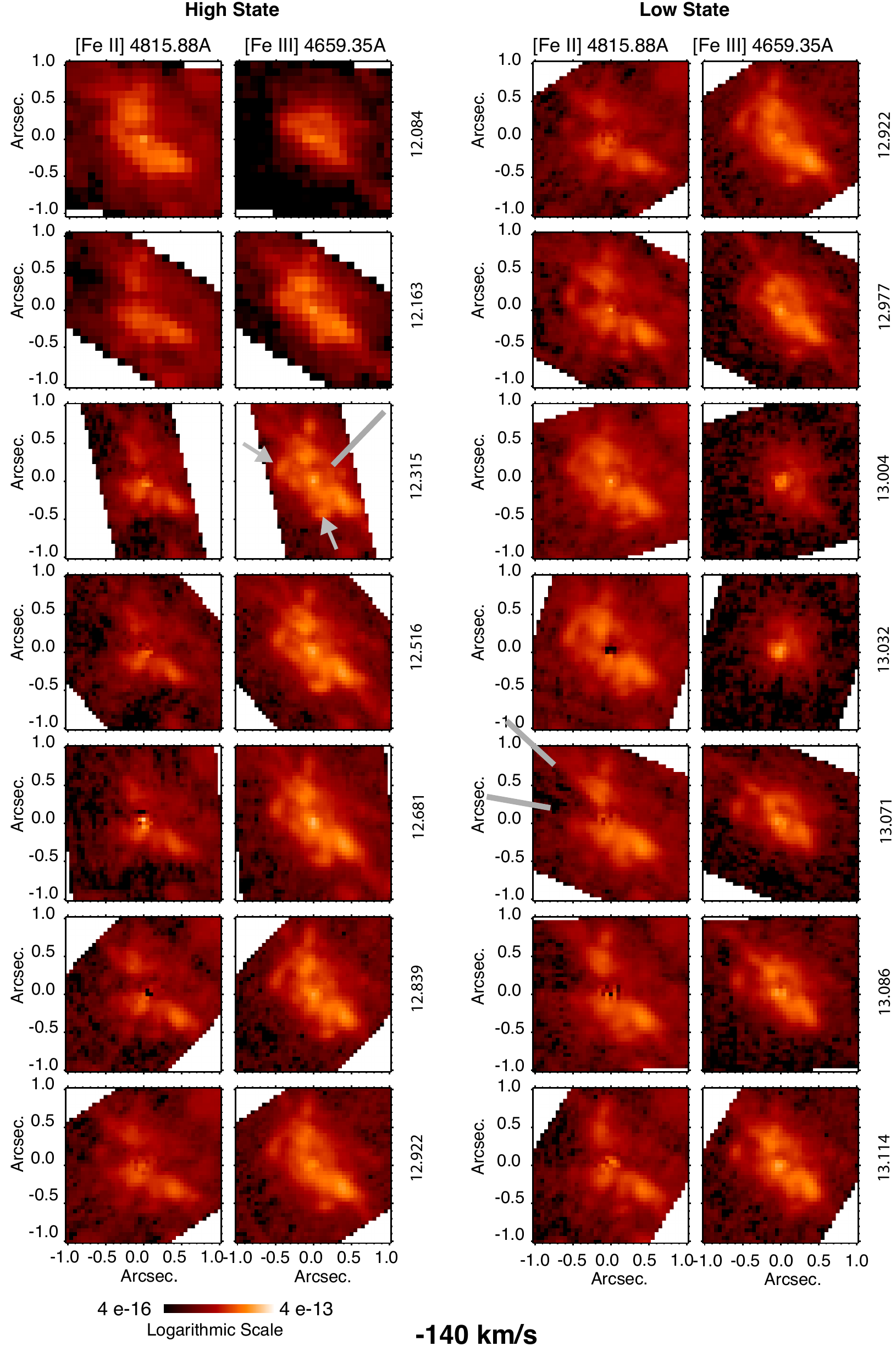}
\caption{{\bf Changes in spatial structure from 2009.4 to 2014.2: $-$140 \kms.} {\bf Left:} Iso-velocity slices of [Fe~II] and [Fe~III] for the seven mappings across the high-ionization state from $\phi=$  12.084 to 12.922.  {\bf Right:} Iso-velocity slices across the low-ionization state from $\phi=$  12.922 to 13.114.
Two [Fe~III] claws (grey arrows, [Fe~III], $\phi=$ 12.315) develop  0\farcs3 radially separated from \ecar\ at PA$=$ 60 and 200\degr\  beginning by $\phi=$ 12.315, radially expand and change in photo-ionization across the cycle. The image of a crab can be imagined, with the body rotated to $-$45\degr\ (axis indicated by grey line).   A dark cone developed after periastron in [Fe~II] (grey lines, [Fe~II], $\phi=$ 13.071), bordered by an [Fe~III] streak.
Note that the very bright, slowly moving structures, including Weigelt C and D, seen in low velocity frames (Fig. \ref{a-40}) have disappeared. 
 \label{a-140}}
\end{figure*}

\begin{figure*}
\includegraphics[height=7.8in,angle=0]{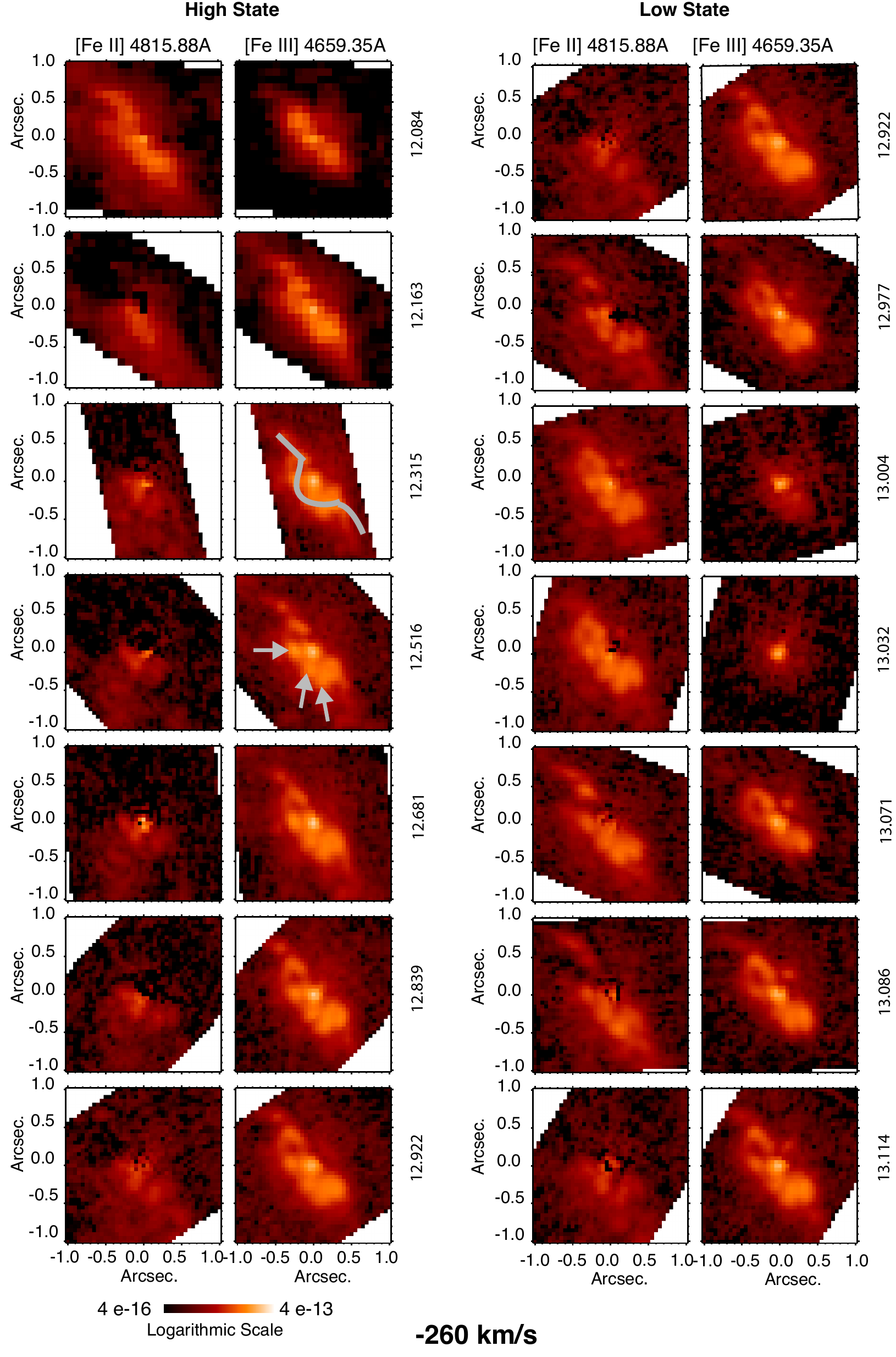}
\caption{{\bf Changes in spatial structure from 2009.4 to 2014.2: $-$260 \kms. Left:} Iso-velocity slices of [Fe~II] and [Fe~III] for the seven mappings across the high-ionization state from $\phi=$ 12.084 to 12.922.    {\bf Right:} Iso-velocity slices across the low-ionization state from $\phi=$  12.922 to 13.114 (see Section \ref{blue} for description). Three arcuate structures, or claws, develop by $\phi=$ 12.315 (grey arrows, [Fe~III] at $\phi=$ 12.516) and persist across the high ionization state. Alternatively, a series of clumps define an $\Omega$-like shape in [Fe~III] by $\phi=$ 12.315 that persists until $\phi=$ 12.681 (grey $\Omega$ in [Fe~III], $\phi=$ 12.315) Then it develops into a figure-eight shape that disappears before periastron, but returns after the low-ionization state ($\phi=$ 13.071 to 13.114). 
\label{a-260}}
\end{figure*}

Considerable variations are identifiable in the blue-shifted structures.  Figs. \ref{a-140} and \ref{a-260} reproduce iso-velocity images for $-$140 and $-$260 \kms\ as examples of these many variations. Several cyclic changes are obvious in the blue-shifted frames: 
\begin{enumerate}
\item \protect{[Fe~II]} is most extensive across the low-ionization state ($\phi=$ 13.004 to 13.071), but fades across the high-ionization state with minimal emission near apastron. Little or no  [Fe~II] is seen at the position of \ecar.
\item \protect {[Fe~III]} is  extensive across the high-ionization state but nearly disappears across the low-ionization state. The peak flux usually at the position of \ecar.
\item With increasing blue-shift, decreasing [Fe~II] emission structures are present, other than across the low-ionization state when [Fe~III] emission fades and is replaced by [Fe~II] emission.
\item \protect{[Fe~III]} emission across the high-ionization state develops into a crab-shaped structure, rotated to $-$45\degr, extending from the northeast to the southwest centered about \ecar. With the two claws that appear at PA$=$ 80\degr, 0\farcs4 and PA$=$ 190\degr, 0\farcs4, the two extended structures PA$=$ 0\degr\ positioned 0\farcs5  and PA$=$ 315\degr\ 0\farcs5, plus the head-like \ecar, the  structure  resembles a crab in Fig. \ref{a-140} at $\phi=$ 12.315.. Appropriately the claws expand outward with orbital phase across the high-ionization state. This structure was noted by \cite{gull09} in the broader velocity interval from $-$200 to $-$400 \kms\ interval.
\item Many [Fe~II] emission structures  lie radially just beyond  [Fe~III] structures relative to  \ecar. 
\item Very little blue-shifted structure is seen to the northwest of \ecar.
\item As noted with the low velocity frames, the maps recorded at $\phi=$ 12.084 and 12.163, allowing for  under sampling in the first observations, are remarkably similar to the maps recorded at $\phi=$ 13.086 and 13.114. 

\end {enumerate}

Specific variations are:
\begin{enumerate}\item The [Fe~III] structure, that appears crab-like complete with body and large claws in Fig. \ref{a-140} at $\phi=$ 12.315,  persists through $\phi=$ 12.977. Most notably the claws expand radially outward with progressing phase. Measures of the [Fe~III] expansion proved to be inconsistent when measured at different phases across the long, high-ionization state. By contrast, measurements of the multiple shells seen in the red-shifted [Fe~II] and [Ni~II] maps led to a consistent expansion velocity of 470 \kms  \citep{Teodoro13}. Careful examination of the multiple velocity frames revealed that the structure shifts tangentially as well as radially. Likely the [Fe~III] emission originates from spatially different regions as \ecar~B slowly changes illumination angle as seen by the fossil shell --analogous to a scanning beam from a (drifting) lighthouse boat illuminating a distant, moving ship through a shifting fog. Here the  fog may be due to thin, compressed primary-wind walls moving in our general direction. Similar,  small- and large-scale effects have been noted previously \citep{Smith04,   madura12a}.

These [Fe~III] claws fade as the high-ionization state shifts towards the low-ionization  state (Fig. \ref{a-140}, $\phi=$ 12.922 to 12.977). Already very faint at $\phi=$ 12.977, the claws, and all other [Fe~III] structures, disappear by $\phi=$ 13.004. Only a diffuse, faint structure remains to the northwest, offset from  core [Fe~III] emission at the position of \ecar. The claw at PA$=$ 80\degr\ reappears by $\phi=$ 13.071 and may be joined by a faint, second claw at PA$=$ 190\degr. Both are about 0\farcs6 separated from \ecar.  

The claws are not seen in [Fe~II] until $\phi=$ 12.922. They brighten to a maximum across the low-ionization state. The PA$=$ 80\degr\ claw disappears by 13.071 but the claw at PA$=$ 190\degr\ may be present until $\phi=$ 13.086. 

\item Shadowing effects become apparent when comparing [Fe~II] to [Fe~III] frames for each phase. An example is to be seen in 
Fig.~\ref{a-140} comparing the forbidden line maps recorded at $\phi=$ 12.163, 13.071, 13.086 and 13.114. An apparent cavity is defined 
between PA$=$ 40\degr\ and 85\degr\ in [Fe~II], but in [Fe~III] a  crab claw appears at PA$=$ 80\degr at $\phi=$ 13.071 while the [Fe~II] persists. A spike of [Fe~III] emission bounds the 
[Fe~II] edge at PA$=$ 40\degr\ at $\phi=$ 12.163 and 13.086. A second spike is seen in $\phi=$ 12.084 and 12.163 extending to the edge of the 
2\arcsec $\times$2\arcsec\ frame at PA$=$ 230\degr, but does not appear one cycle later at $\phi=$ 13.086 nor 13.114. These spikes are NOT instrumental as they do not conform to a row or column of 
the maps and are seen at the same PAs even though the \hst-defined PA for the \stis\ aperture changes 
between visits. A second spike can be seen in [Fe~III] at phases $\phi=$ 12.516 and 12.681 at PA$=$ 30\degr. This spike also appears 
in [Fe~III] at $\phi=$ 12.977 along with the spike at PA$=$ 225\degr.

These shadows and spikes, seen at $\phi=$ 13.071 to 13.114, most likely trace out boundaries in wind-wind structures and fossil structures that pass or block the FUV radiation, thus leading to gaps in [Fe~III] high-velocity emission that are filled in by [Fe~II] emission. A second, much less obvious, shadow may be to the south of \ecar, in the direction from PA$=$ 160 to 190\degr. The region between the two shadows, PA$=$ 90 to 160\degr, shows up in [Fe~II]. 

\item The  $-$260 \kms\ [Fe~III] iso-velocity images in Fig. \ref{a-260}\ reveal three claws: 0\farcs2 at 80\degr\ and 180\degr\ plus 0\farcs4 at 200\degr. By $\phi=$ 12.977, these claws show up in [Fe~II], persist in [Fe~II] through $\phi=$ 13.086, then begin to fade by $\phi=$ 13.114.

By contrast the [Fe~III] structure, seen as a Crab-like structure from $\phi=$ 12.315 to  12.922 in $-$140 \kms\ iso-velocity images (Fig. \ref{a-140})  distorts into an $\Omega$-shape, rotated by 135\degr (Fig. \ref{a-260}). With advancing phase, the [Fe~III]  shifts in form to a near-figure-eight at $\phi=$ 12.977,  shrinks to  emission at the position of \ecar\ plus faint,  diffuse emission to the south and west, then re-emerges as the figure-eight  by $\phi=$ 13.071. The [Fe~III] again appears as a distorted $\Omega$\   by $\phi=$ 13.114. 

In both $\phi=$ 12.084 and 13.086 at $-$260 \kms\ (Fig. \ref{a-260}), a remarkable arc pops up in [Fe~II] pointing to the northeast, but with an apparent shadow cutting across from east to west about 0\farcs1 north of \ecar. At $\phi=$ 13.086, this [Fe~II] arc appears to be  shadowed before PA$=$ 45  and after PA$=$ 22\degr. The apparent shadows may indicate regions of higher photoionization adjacent the [Fe~II] arc.
\end{enumerate}

\subsubsection{The red-shifted structures}\label{red}

\begin{figure*}
\includegraphics[height=7.8in,angle=0]{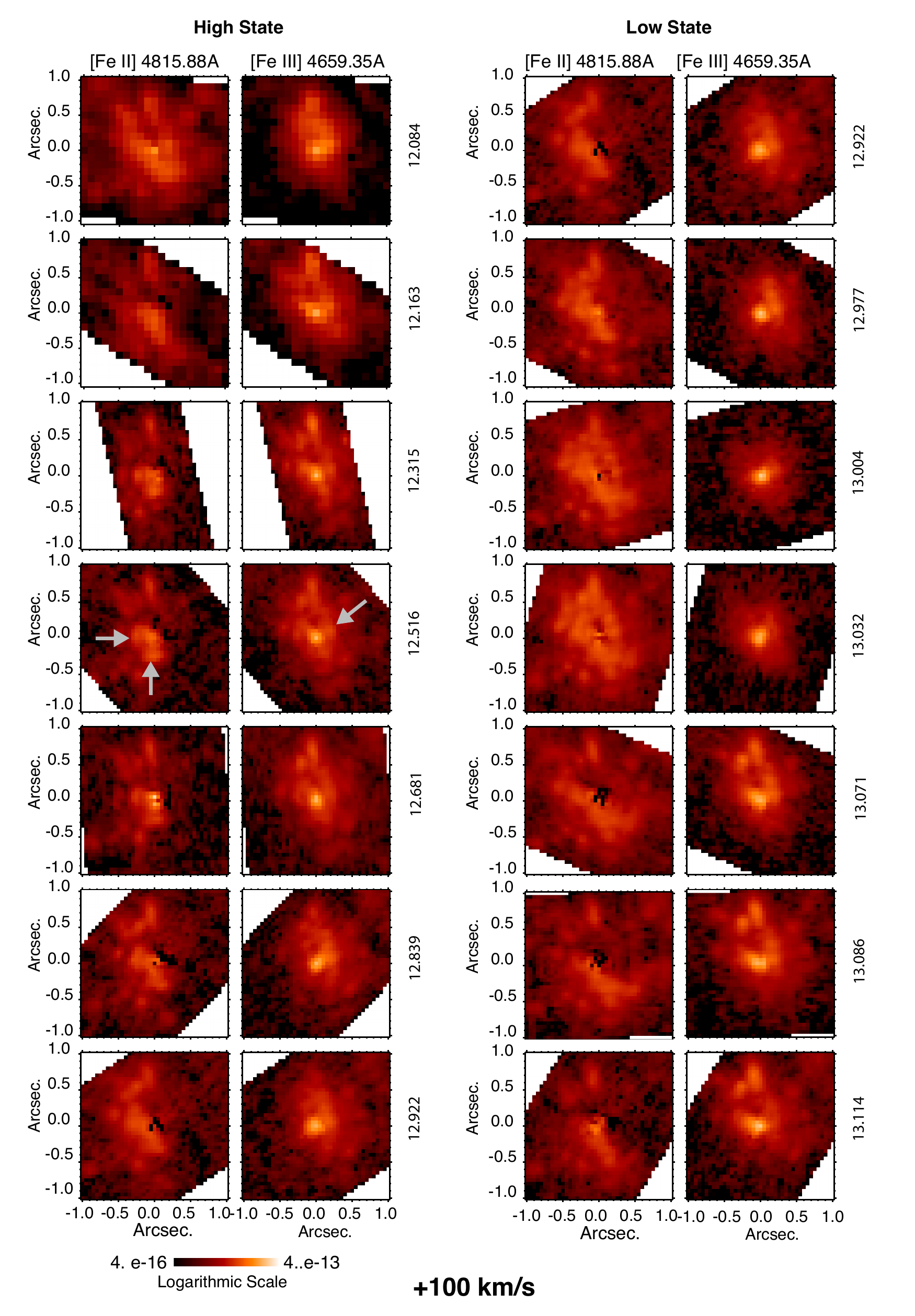}
\caption{{\bf Changes in spatial structure from 2009.4 to 2014.2: $+$100 \kms.} {\bf Left:} Velocity slices of [Fe~II] and [Fe~III] for the seven mappings across the  high-ionization state from phases $\phi=$ 12.084 to 12.922.   {\bf Right:} Seven mappings across the low-ionizaton state from phases $\phi=$ 12.922 to 13.114. Red-shifted [Fe~II] structures, situated on the far side of \ecar\ are visible, despite the intervening, foreground fossil winds. Two [Fe~II] arcuate structures develop by $\phi=$ 12.315 (grey arrows, [Fe~II] at $\phi=$  12.516) and a [Fe~III] arcuate structure at PA$=$ 280\degr\ (grey arrow, [Fe~III] at $\phi=$ 12.516) that form a nearly complete ring. Multiple clumps appear in [Fe~II] during the low-ionization state ($\phi=$ 13.004 to 13.032).
The Weigelt objects and hook structure, seen in low velocity (Fig. \ref{a-40}), are no longer present. 
\label{a+100}}
\end{figure*}

\begin{figure*}
\includegraphics[height=7.8in,angle=0]{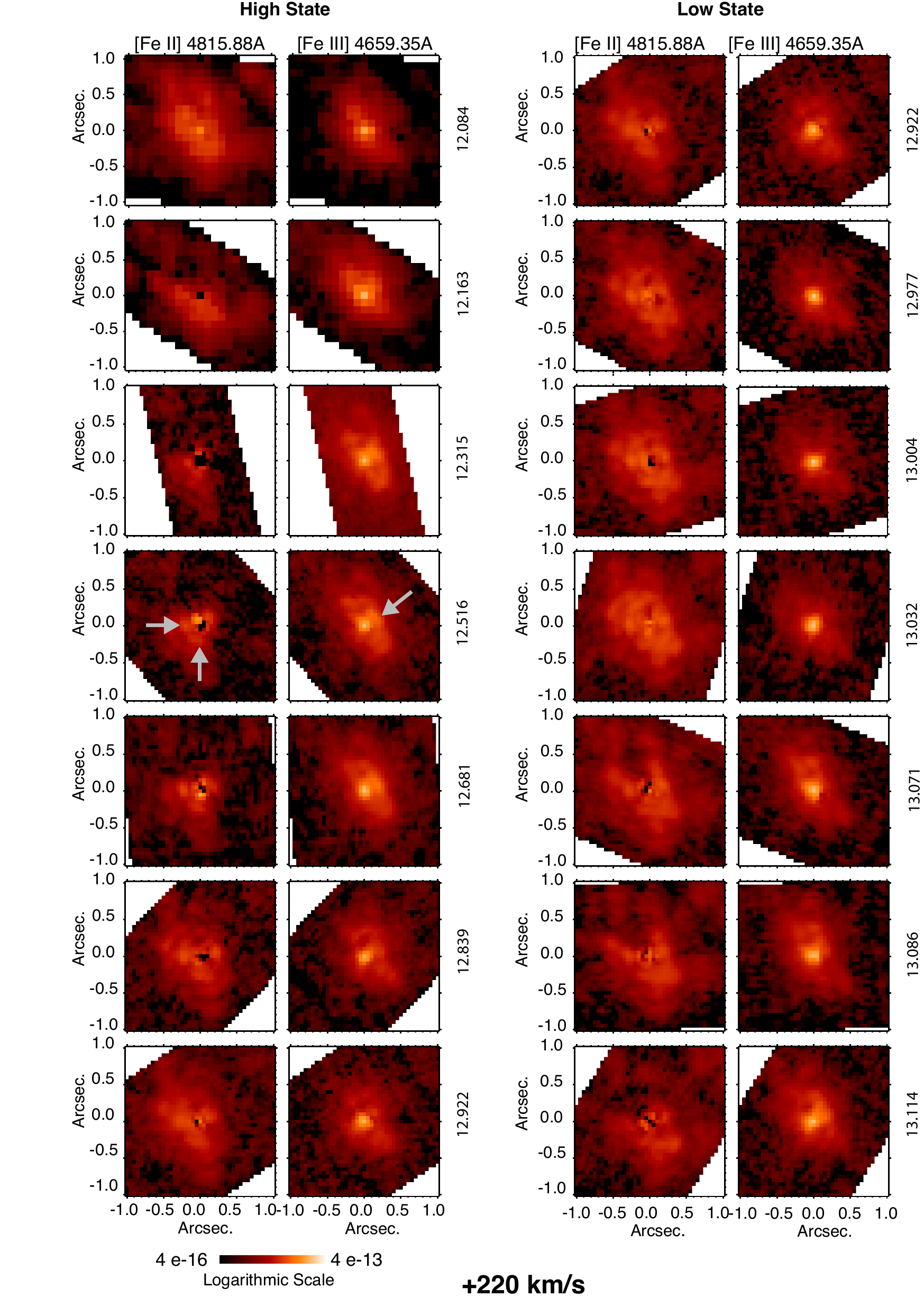}
\caption{{\bf Changes in spatial structure from 2009.4 to 2014.2: $+$220 \kms. Left:} Velocity slices of [Fe~II] and [Fe~III] for the seven mappings across the high-ionization state from phases $\phi=$ 12.084 to 12.922.   {\bf Right:} Seven mappings across the low-ionization state from phases $\phi=$ 12.922 to 13.114. The [Fe~II] emission is confined to the southeast quadrant across the last half of the  high-ionization state ($\phi=$ 12.516 to 12.922). However across the low-ionization state, two large, continuous arcs form from north counter-clockwise to south, outlining large portions of the compressed primary wind. As seen in Fig. \ref{a+100}, two [Fe~II] arcuate structures develop by $\phi=$ 12.315 (grey arrows, [Fe~II] at $\phi=$  12.516) and a [Fe~III] arcuate structure at PA$=$ 280\degr\ (grey arrow, [Fe~III] at $\phi=$ 12.516) that form a nearly complete ring.A noticeable [Fe~III] arc nearly completes the first shell ring  formed in [Fe~II], 0\farcs2 distance from \ecar, from $\phi=$ 12.516 to 12.922.
\label{a+220}}
\end{figure*}

The real surprise was revealed in maps across the low-ionization state from $\phi=$ 12.977 to 13.032 (Figs. \ref{a+100} and  \ref{a+220})\footnote {These iso-velocity slices do not include contamination   from the weak [Fe~II] or Fe~II lines that contribute significantly above $+$300 \kms. See Appendix \ref{reduce}.}. Short, arcuate structures appear in [Fe~II]  extending from the central region as far out as 0\farcs9! As the FUV radiation disappeared from regions where iron was doubly-ionized across the high-ionization state, additional, unanticipated [Fe~II] structures suddenly appeared (in Figs. \ref{a+100} and \ref{a+220}, see the [Fe~II] frames across $\phi=$13.004 to 13.086 compared to the [Fe~II] frames across $\phi=$12.315 to 12.922) . These [Fe~II] structures are more extensive, i.e. are found in more distant regions, since the critical density is significantly lower than for [Fe~III].

\begin{enumerate}
\item The innermost, [Fe~II] red-shifted arcuate-structures, seen at $\phi=$ 12.315 to be  0\farcs2 radially separated from \ecar\ at PA$=$ 80\degr\ and 180\degr,  expand at 470 \kms\ across the high-ionization state \citep{Teodoro13}. These innermost arcs are readily apparent   through $\phi=$ 12.977, but become very diffuse from $\phi=$ 13.004 to 13.114, following the shapes of the earlier, innermost structures seen one cycle earlier $\phi=$ 12.084 and 12.163.
\item Three sets of  [Fe~II]  arcuate structures,  0\farcs2, 0\farcs45 and 0\farcs7 separation from \ecar\ at PA$=$ 80\degr\  and 180\degr\ were visible across the high-ionization state \citep{Teodoro13}, but additional small arcuate structures are present in [Fe~III]  to the northwest of \ecar\ from $\phi=$  12.315 to 12.922. The innermost set of [Fe~II] and [Fe~III] arcuate structures at $\phi=$ 12.315 to 13.839 form a nearly complete ring in figs. \ref{a+100} and \ref{a+220}.
\item Many more [Fe~II]  arcuate structures appear at greater separations from \ecar\ across the low-ionization state ($\phi=$ 13.004 to 13.071).

\item Some of the additional [Fe~II] structures, allowing for radial expansion, track with structures in [Fe~III] across the high-ionization state from $\phi=$ 12.315 to 12.922.

\item The two inner structures are nearly complete rings in [Fe~II] during the low-ionization state ($\phi=$ 13.004 to 13.071) 
\item Two elongated structures are present in the $+$100 \kms\ frames  at $\phi=$ 13.071 through 13.114 in Fig. \ref{a+100}. The north-pointing structure is visible throughout the cycle in [Fe~II] and appears in [Fe~III] through $\phi=$ 12.681, but disappears by $\phi=$ 12.839, returning by $\phi=$ 13.071. The northeast-pointing structure is visible in [Fe~II] at 40\degr\ from $\phi=$ 12.084 to 12.839 but fades late in the high state and may be returning by $\phi=$ 13.114.

\item A third arc to the west-northwest at 0\farcs2 can be traced from $\phi=$ 12.084 to 12.977 at $+$100 \kms\, but disappears by $\phi=$ 13.004. It is also visible in the $+$220 \kms\ frames from $\phi=$ 12.315 to 12.681 (Fig. \ref{a+220}).

\item The arcuate structures in [Fe~II] shrink with increasing velocity and converge, projected onto  \ecar's position, by $+$400 \kms (See Fig. \ref{FeIImosaic} for complete coverage of these structures across all phases and red-shifted velocities). Models of the southeast quadrant shells suggest nested shells that were nearly spherical in shape \citep{Teodoro13}. 
\end{enumerate}

\subsection{Difference mapping to show  the photo-ionization and expansion effects across the high-ionization state}\label{expandlight}

For most of the binary period, the winds of \ecar~B push the winds of \ecar~A aside forming a large cavity in the current wind-wind collision structures. The current wind interaction structures, not reseolvable by \hst\ have now resolved by {\it VLTI}\ (Weigelt et al. submitted). The long-term accumulation from many cycles leads to an extensive cavity that projects northwest of the binary. The FUV radiation from \ecar~B escapes and  ionizes much of the gas within the cavity and the outer portions of the compressed, modulated wind of \ecar~A. Evidence of this asymmetry is demonstrated by the several arcseconds offset of the H~II region from \ecar\ as imaged at radio wavelengths by \cite{duncan03}. 

As each periastron approaches, \ecar~B dips deeply into the primary wind that  absorbs most of the FUV radiation. Most of the [Fe~III] emission structures disappear. \ecar~B penetrates so deeply that it carves a tunnel-like cavity within the primary wind, then pushes and photo-ionizes the primary wind until the extensive cavity is expanded even further. The tunnel continues to expand above and below the orbital plane creating a compressed shell that persists for multiple cycles \citep{madura10, Gull11, Madura12, madura13, Clementel14, Clementel15a, Clementel15b}. 

Multiple clumps appear in [Fe~II], located to the southeast at red-shifted velocities (Figs. \ref{a+100} and \ref{a+220}). Where FUV radiation impinges upon these compressed shells at, or near critical densities of $n_e=10^7 cm^{-3}$, [Fe~III] 4659.35\AA\ is emitted. Elsewhere in the expanding shells, mid-UV from the primary star, and reprocessed FUV radiation contributed from the secondary star, leads to singly ionized iron. At densities close to  $n_e=10^6 cm^{-3}$, [Fe~II] 4815.88\AA\ is emitted. We use these emissions to trace clumpy arcs of  compressed primary wind and to determine how these winds evolve across multiple 5.54-year periods.

One means to display these changes in  wind structure is by normalized differencing of maps by using:
$Change = (M_P-M_A)/(M_P+M_A)$, where M$_P$ is a 2-D map  near periastron, and M$_A$ is a 2-D map  near apastron. We provide iso-velocity slices of normalized differences in Figs. \ref{shells} and \ref{claws}. Note that noise in the extreme velocity iso-velocity images displayed in Figs. \ref{shells} and \ref{claws} is significantly larger due to vey low signal.

\subsubsection{Difference maps in [Fe~II]} \label{expand}

\begin{figure*}
\includegraphics[width=6.9in]{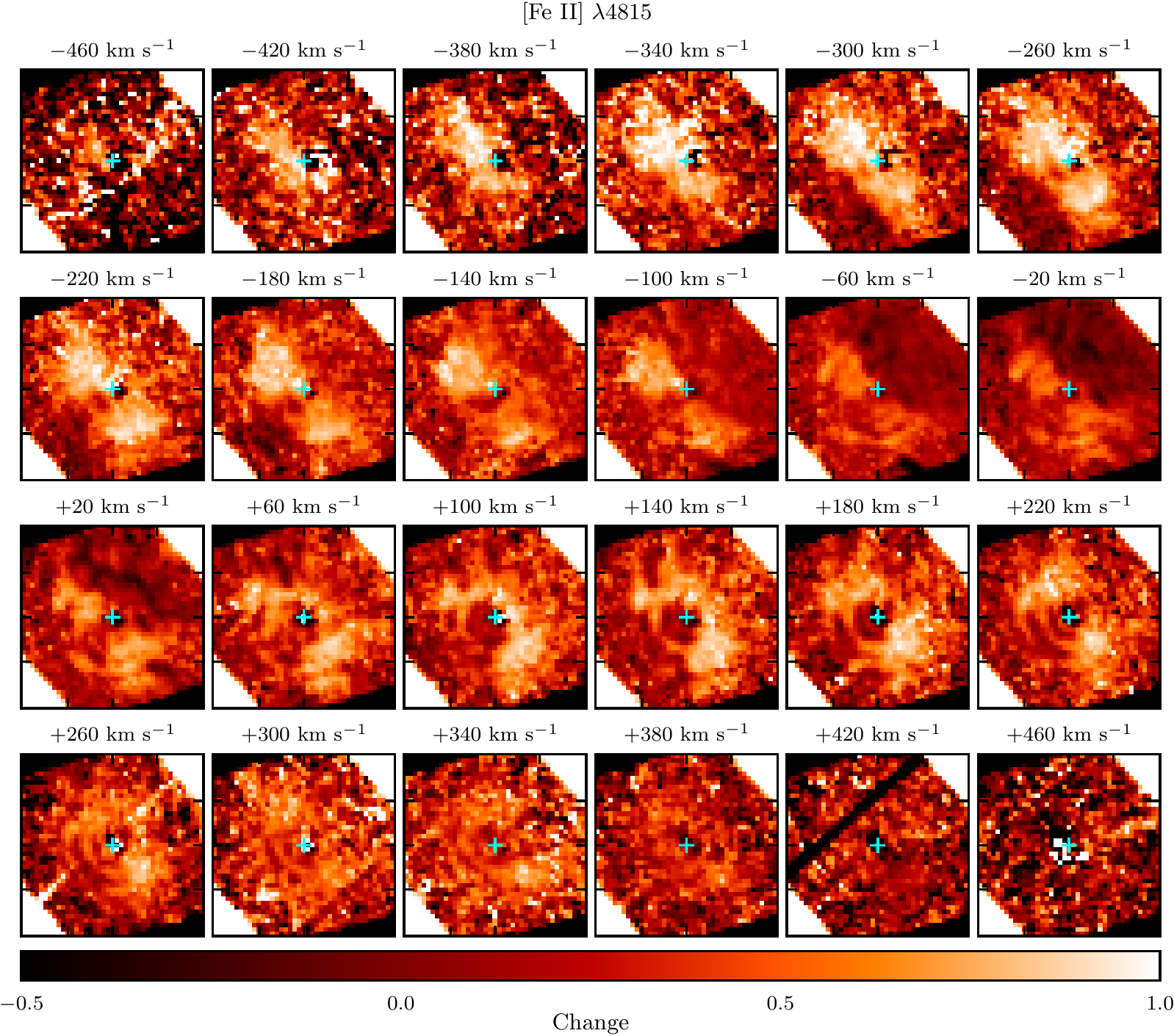}
\caption{{\bf Expansion of the [Fe~II] fossil shells between $\phi=$ 12.516 to 13.004 (2.7 years across the high-ionization state) and the light house effect.} Displayed is the normalized difference between the fluxes of  [Fe~III] recorded at $\phi=$ 12.516, well into the high-ionization state, and at $\phi=$ 13.004, a half cycle later, just as the low-ionization state begins. Each frame is a 40 \kms-wide slice of the [Fe~III] profile starting at $-$380 \kms\ in the upper left corner and ending at $+$380 \kms\ in the lower right corner. White indicates regions where [Fe~II] has increased.  
 \label{shells}}
\end{figure*}
Related changes in structure occur in [Fe~II] from apastron to periastron (Fig. \ref{shells}, see also \cite{Teodoro13}). The overall effect in the differencing images shows up as brightening (white) of [Fe~II] at periastron relative to apastron. In the radial direction from \ecar\ towards these red-shifted shells, the  FUV is blocked by the primary wind in the current wind-wind interaction structures. Only the MUV can escape leading to singly-ionized iron. 

Four large scale changes can be seen in the narrow velocity images:
\begin{enumerate}

\item $-$100 to $+$340 \kms: Structures  of alternating bright/dark arcs appear counterclockwise from PA$=$ 90 to 180\degr. These arcs  are the remnants of  compressed primary-wind shells caused by the passage of \ecar~B through the massive wind of \ecar~A during periastron as analyzed by \cite{Teodoro13}.

\item $-$380 to $-$100  \kms: Diffuse structures extend counterclockwise from PA$=$ 45 to 90\degr and 180 to 220\degr, slightly offset to the southeast from \ecar. These correlate with [Fe~III]-bright structures at apastron and represent the outer, chaotic regions of the extensive wind-wind cavity.

\item $+$100 to $+$340 \kms: Diffuse structures extend clockwise to the northwest, with an arcuate boundary grazing the position of  \ecar. These structures also correlate with extended [Fe~III]-bright structures at apastron and again are associable with outer, chaotic regions. 

\item $-$100 to $+$20 \kms: A lack of [Fe~II] emission to the northwest that defines a large arcuate cavity with central edge grazing the position of \ecar. This darkened region follows the shape of the slow-moving structures as illustrated in Fig. \ref{a-40}. Additionally the shadows due to the `ears' or filaments to the north at PA$=$ 35\degr\  are described subsection \ref{slow}.
\end{enumerate}

\subsubsection{Difference maps in [Fe~III]} \label{light}
\begin{figure*}
\includegraphics[width=6.9in,angle=0]{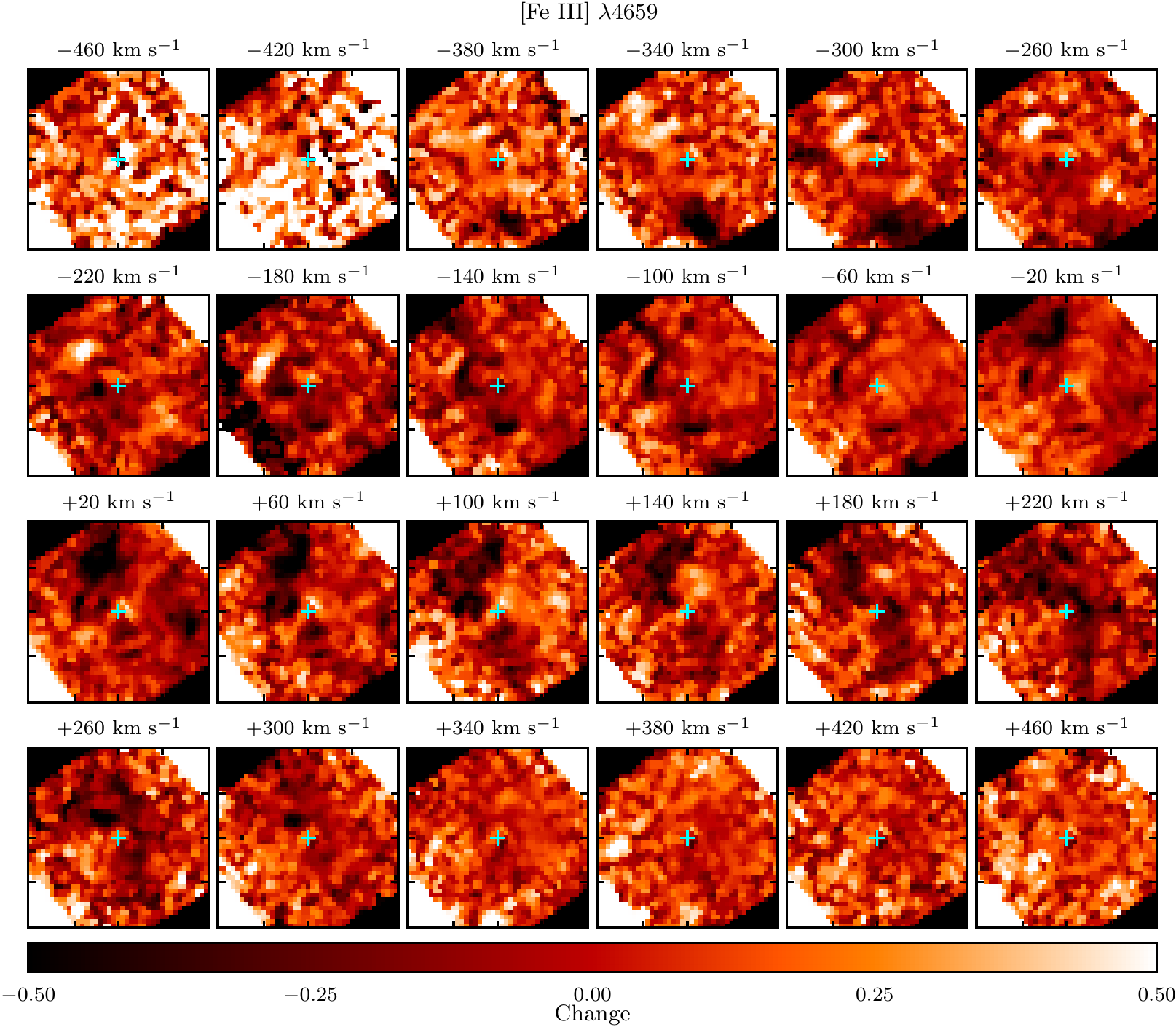}
\caption{{\bf Shell expansion and the lighthouse effect of  [Fe~III] emission between $\phi=$ 12.516 to 12.922.} 
 Each frame is a 40 \kms-wide slice of the normalized difference starting at $-$380 \kms\ in the upper left corner and ending at $+$380 \kms\ in the lower right corner.  [Fe~III] intensity has dropped in regions displayed as black, and has increased in regions showing white.  
 Expansion is revealed by paired white and black arcs to the northeast and southwest.
 The lighthouse effect, due to obstructing clouds deep within the structures that block the FUV, is especially prominent in the $-$60 to $+$60 \kms\ frames. Structures about 0\farcs6 north by northwest darken while structures 0\farcs3 due north relative to \ecar\ brighten in the $-$100 to $+$100 \kms velocity range.
 \label{claws}}
\end{figure*}

Changes in [Fe~III] across the long, high-ionization state are represented in Fig. \ref{claws}, which is a normalized difference between phases $\phi=$ 12.516 (apastron) and 12.922 (the last  mapping before periastron when extensive [Fe~III] is present). Each iso-velocity image is a 40 \kms\ velocity slice for the intervals from $-$380 to $+$380 \kms.

Most noticeable are two effects:
\begin {enumerate}
\item Two white arcs, one 0\farcs5 at PA$=$ 45\degr\ and the other 0\farcs5 that PA$=$ 220\degr, appear in the $-$340  through $+$20 \kms\ iso-velocity images in Fig. \ref{claws}. At $-$180 \kms\ less-defined interior dark pair of arcs appears signifying expansion of the [Fe~III] structures across the 2.7-year interval between apastron and pre-periastron. These paired arcs can be followed to $+$200 \kms. A fainter, bright  pair   develops at velocities red-ward of $-$20 \kms. These arcs are Fe$^{++}$ portions of the compressed primary wind that continue to expand outward and apparently persist two cycles (eleven years) after being formed!

\item 
Irregular structures 0\farcs6 centered on PA$=$ 10\degr\ first appear as two, thin, dark streaks by $-$100 \kms, become diffuse with red-shifting velocity until disappearance by $+$220 \kms.   The darkening shifts clockwise to the east with increasing positive velocity and can be traced to $+$180 kms. The apparent fading and shifting positions of [Fe~III] emission are due to clumps of material deep within the complex structure that block the FUV radiation from \ecar~B as it slowly moves across apastron. Hence the FUV radiation scans clockwise across distant structures.
\end{enumerate}

\section{Discussion} \label{results}

 The  [Fe~III] and [Fe~II] structures paint a consistent picture of a series of partial fossil-wind shells with a large cavity, built up over multiple cycles, but which also contains obscuring,  slow-moving, massive clumps of material originating from the two ejections during the 1840s and 1890s. 
 
 The [Fe~III] maps reveal the highly-ionized ($>$16.2 eV) portions of the fossil winds (Fig. \ref{FeIIImosaic}). Across the high-ionization state, these structures extend counterclockwise from the northeast to the southwest with blue-shifts reaching to $-$450 \kms\ (Figs.\ref{a-140}, \ref{a-260}, and \ref{FeIIImosaic}). Much fainter, somewhat red-shifted [Fe~III] structures extend short distances to the northwest (Figs. \ref{FeIIImosaic}, \ref{a-40} and \ref{a+100}). All [Fe~III] structures disappear across the low-ionization state except for faint emission at the position of \ecar\ (within the current cavity enclosed by the primary wind) and very faint, diffuse emission within 0\farcs3 to the northwest (small breakouts in the primary wind). 
 
 The [Fe~II] maps reveal the low-ionized ($<$7.9 eV) portions of the fossil winds (Fig. \ref{FeIImosaic}). Across the high-ionization state, relatively faint, blue-shifted [Fe~II] structures, that are  more diffuse than those seen in [Fe~III], lay radially just beyond the [Fe~III] structures (Figs. \ref{FeIImosaic}, \ref{a-140} and \ref{a-260}). At red-shifted velocities, multiple arcs appear in the [Fe~II]  (Figs. \ref{FeIImosaic}, \ref{a+100} and \ref{a+220}).  As analyzed by \cite{Teodoro13}, the arcs seen in the southeast across the multi-year high-ionization state were measured to expand as nearly spherical shells at 470 \kms.  The terminal wind velocity  of \ecar~A was determined to be \vinf$_{,A}=$ 420 \kms\ \citep{Groh12}. Modeling of the interacting winds by \cite{madura13} showed that the secondary wind interacting with the primary wind could accelerate these shells to 470 \kms, consistent with the measured velocities of these fossil shells. The net effect would be variable expansion velocity of different portions of each shell dependent upon the total momentum imparted on the primary wind by the colliding secondary wind.
 
 Across the high-ionization state, many [Fe~II]  arcs are visible in the  southeast quadrant, but across the relatively brief low-ionization state, even more [Fe~II] arcs appear surrounding \ecar\ (Figs. \ref{FeIImosaic}, \ref{a+100} and \ref{a+220}). Three, possibly four, shells appear. Re-examination of the [Fe~III] velocity maps confirm that some of the inner arcs are  nearly complete, interior shells at low red-shifted velocities as a complementary portion appears in [Fe~III] across the high-ionization state.   Additional arcs and clumps, likely from shells even older than the relatively complete shells detected here, appear in [Fe~II] across the low-ionization state at greater distances from \ecar, likely due to relatively lower critical density for the transition. 
 
 The presence of these multiple [Fe~II] shells, demonstrated by the multiple pairs of arcuate structures (Figs. \ref{a+100} and \ref{a+220}) plus the nearly complete arcs in the normalized differences of (Fig. \ref{shells}) strongly reinforce that the momentum balance of the winds has not changed substantially over the past several 5.5-year cycles. The periodic spacing of the arcuate structures, as measured by \cite{Teodoro13}, indicate near-uniform spacing. We suggest a  limit of twenty percent to changes in the momentum balance. Additional observations should be made for confirmation.

 The [Fe~III] and [Fe~II] mappings describe structures that are highly ionized on the near side and partially ionized on the far side of \ecar. Moreover the blue-shifted structures seen in  [Fe~III] are much more distorted and chaotic than the red-shifted structures seen in [Fe~II]. Such can only be if the FUV source spends most of the orbit on the near side, and the same source provides the less-massive, faster wind. Hence, \ecar~B, across the long, high-ionizaton state, centered upon apstron, the FUV and much faster, less massive wind source must be on the near side of \ecar~A and the rapid passage through the massive wind of \ecar~A by \ecar~B must occur on the far side. This is contrary to the orbit proposed by \cite{abraham05b} and \cite{Kashi15}, but consistent with models suggested by \cite {Pittard02}, \cite{Nielsen07} and \cite{madura10}. 
 
The iso-velocity images centered at $+$420 to $+$460 and $-$420 to $-$460 \kms\ show little emission (Figs. \ref{FeIIImosaic} and \ref{FeIImosaic}), since these intervals include the terminal velocity of the primary wind, \vinf$_A=$ 420 \kms\ \citep{Groh12} and the partially-accelerated shell velocity, 470 \kms\  \citep{Teodoro13}. The [Fe~III] iso-velocity images for $-$420 \kms\ show very weak diffuse emission extending to the northeast, bounded by PA$=$ 30 to 70\degr\ early and late in  the high-ionization state (Fig. \ref{FeIIImosaic}). The [Fe~II] iso-intensity images show the same structure between $\phi=$ 13.004 and 13.032 (Fig. \ref{FeIImosaic}). A noticeable difference is in the [Fe~II] frames for $\phi=$ 12.084 and 12.163 as there appears to be some weak [Fe~II] emission not seen in the high velocity frames for $\phi=$ 13.086 and 13.114.

Recently, \cite{Abraham14} observed \ecar\ at several hydrogen transitions, H n$\alpha$,  with the Atacama Large Millimeter/submillimeter Array ({\it ALMA}). They postulated that the structures seen in selected  H I 21$\alpha$\  to 42$\alpha$\ lines with $\approx$1\arcsec\ angular resolution were due to the formation of a baby Homunculus in the 1940's timeframe. These structures are quite similar in nature to 3-cm, radio-continuum structures imaged by \cite{duncan03} with the Australian Telescope Compact Array ({\it ATCA}). Both the continuum and the recombination-line structures appear to be offset to the northwest of the [Fe~II] structures identified in the present paper. 

We posit that these structures are the ionized-complement in the  large cavity blown out over many cycles by the secondary wind and photo-ionized by the FUV radiation from the secondary star. As noted by \cite{duncan03}, the radio continuum, while centered on \ecar, extends to the northwest growing and shrinking with binary phase over the 5.5-year period. The [Fe~III] structures mapped across the most recent cycle grow and shrink towards the northwest, but are confined to closer regions than the radio continuum. Likely this is because the [Fe~III] emission reveals the relatively high-density regions (n$_e=$  10$^7$cm$^{-3}$) while the radio continuum originates from much lower, but larger regions in the extended wind cavity.

The recent observations of \ecar\ with the {\it VLTI}  (Weigelt et al. submitted) of the H I Br$\gamma$\ line (angular resolution 6 mas, R $= \lambda/\delta\lambda=$12,000) directly image the current wind-wind interactions. The {\it VLTI} iso-velocity imagery  supports  the fossil wind structures discussed herein.

 The Weigelt objects and related, slowly-moving structures, as suggested by their radial velocity (Fig. \ref{a-40}), are noticeable only on the near side. The regularly spaced arcs on the far side would not exist were there clumps moving at low velocities ($+$10 to $+$100 \kms). The existence of the regularly-spaced, slow-moving, red-shifted arcs suggests that no analogs to the Weigelt blobs and extended structures exist on the far side. 
Does this suggest that the 1890s ejection occurred during periastron passage, leading to rapidly moving ejecta on the far side of \ecar, but that only some, very slow-moving ejecta flew out in our direction on the opposite side of the encounter? Is this another clue about the mechanism?

Curiously the detected clumps, of which the Weigelt objects are the brightest, are on the blue-shifted, highly-ionized side of the system. Based upon what is seen in [Fe~II], the red-shifted side of \ecar\ appears to be a fossil wind that is flowing uniformly over the past several 5.54-year periods with shell-like structures carved out  by the rapid  passage of \ecar~B  during each periastron event. 
No red-shifted, slowly moving clumps appear to be present. 

\section{Conclusions}
\label{conclude}

A series of observations has been accomplished across one 5.54-year binary period. Using \stis\ on  \hst, we have systematically mapped the central region surrounding \ecar\ with 0\farcs13 angular resolution and R$=\lambda/\delta\lambda=$8,000. Focusing on spatial- and  velocity-differentiated structures seen in the light of [Fe~III] 4659.35\AA\ and [Fe~II] 4815.88\AA, we have followed the changes in photo-ionization and expansion of multiple structures surrounding \ecar. We find these structures to be fossil remnants of wind-wind interactions remaining from periastron passages 5.5, 11.1, 16.6 and possibly 22.2 years previously. 

Photo-ionization changes of the fossil structures are brought on by modulation of the FUV radiation that originates from \ecar~B. The  companions spend most of the orbit  at  30 AU separation ($=$1.3 mas at a distance of 2300 pc). 
For over five years of each binary period, much of the winds are highly-ionized and glow in [Fe~III]. However across periastron, when the stars approach as close as 1.5 AU,  \ecar~B plunges deeply into the very massive, optically-thick, primary wind. For a few months, the FUV radiation from \ecar~B is blocked by the primary wind.  The wind structures become less-ionized leading to [Fe~II] emission originating from much of the extended wind structures. 

While the secondary wind interacts with the primary wind throughout the entire binary orbit, the secondary star passage deeply through the primary wind  produces a tightly wound tunnel across each periastron, resulting in an expanding, compressed shell. Given the thermal velocities of the primary wind being much smaller than its terminal velocity, the compressed shell persists and expands outward from the binary. Both stars have such massive winds that, for several 5.5-year cycles, the densities of these compressions exceed  n$_e\approx$10$^7$cm$^{-3}$, leading to forbidden emissions from Fe$^{++}$ and Fe$^+$ originating in portions of the shells. Hence we are able to trace the persistent fossil shells.

The fossil shells persist only for a fraction of the region surrounding \ecar, namely the far, red-shifted side where the periastron event occurs (Figs. \ref{FeIImosaic}, \ref{a+100} and \ref{a+220}). The hot, secondary star spends most of the orbit on the near side of \ecar~A, blowing an ionized, low-density cavity across each high-ionization state. Both the 3-D hydrodynamical models of \cite{Okazaki08, parkin09, parkin11, Madura12,madura13, Clementel14, Clementel15a} and the present observations support the concept that the  cavity of the current wind-wind interaction eventually intersects with previous ionized cavities, building  an ever larger structure that extends to the northwest of \ecar\ (Fig. \ref{shells}). The blue-shifted [Fe~III] structures extending from the northeast to the southwest (Figs. \ref{FeIIImosaic}, \ref{a-140} and \ref{a-260}) define the boundaries of this cavity in the form of highly-ionized primary wind structures.

The region between the well-defined [Fe~II] shells in the red-shifted southwest and the blue-shifted [Fe~III] structures include a more chaotic region. This is a transition region between the cavity blown by \ecar~B and the modulated shells from the wind of \ecar~A. In this region and within the wind-blown, large cavity, exist slow-moving clumps of material. The brightest of these are the Weigelt objects, C and D. Their proper motion and position suggests that they were ejected by the system in the 1890s event \citep{weigelt12}. 

The persistence of the multiple shells in [Fe~II] indicates:  
\begin {enumerate}
\item  There are currently no large, slowly moving clumps of gas, like the Weigelt objects, on the far side of \ecar.  This may be an important clue to the major ejection event of the 1890s and possibly the 1840s. Was the event such that the faster moving ejecta was in all directions except towards apastron? Could the explosive event have been at or near periastron leading to an asymmetrical ejection not only in the polar regions but in the orbital plane? 

\item The  mass loss momentum and energy balance of the binary system has not changed significantly over the past several cycles. 
Does this indicate that \ecar\ has settled down for the near term? Could this indicate that the events of the 1840s and the 1890s were not pre-supernova events, but rather were the result of a stellar merger, or some other mechanism?

\end{enumerate}
We leave the conclusion to the modelers and to future observations.
%
%
\section*{Acknowledgements}
%
%
%

TIM was supported by the NASA Postdoctoral Fellowship Program. TRG, TIM and MT received support from STScI grants 12013, 12750, 12508 and 13054 through June 2015. TRG also thanks Gerd Weigelt and the Max Planck Institute for Radioastronomy for  delightful stays in the fall of 2015 and spring of 2016. AD acknowledges the continuing financial support from FAPESP. AFJM is grateful for financial support from NSERC (Canada) and FQRNT (Quebec). NDR acknowledges postdoctoral support by the University of Toledo and by the Helen Luedtke Brooks endowed Professorship. We gratefully thank Ms. Beth Perriello (STScI) for the extraordinary support scheduling visits at critical intervals that led to the success of this very challenging series of observing programs. We thank anonymous referee for helpful editorial comments. And most importantly, we thank the  Eta Car Bunch, a truly innovative group extended across many locations but focused on one massive binary!

\bibliographystyle{mnras}
\bibliography{refs.bib}

%
\appendix
%
\section{Data Reduction}
\label{reduce}

\begin{figure}
\includegraphics[width=\columnwidth]{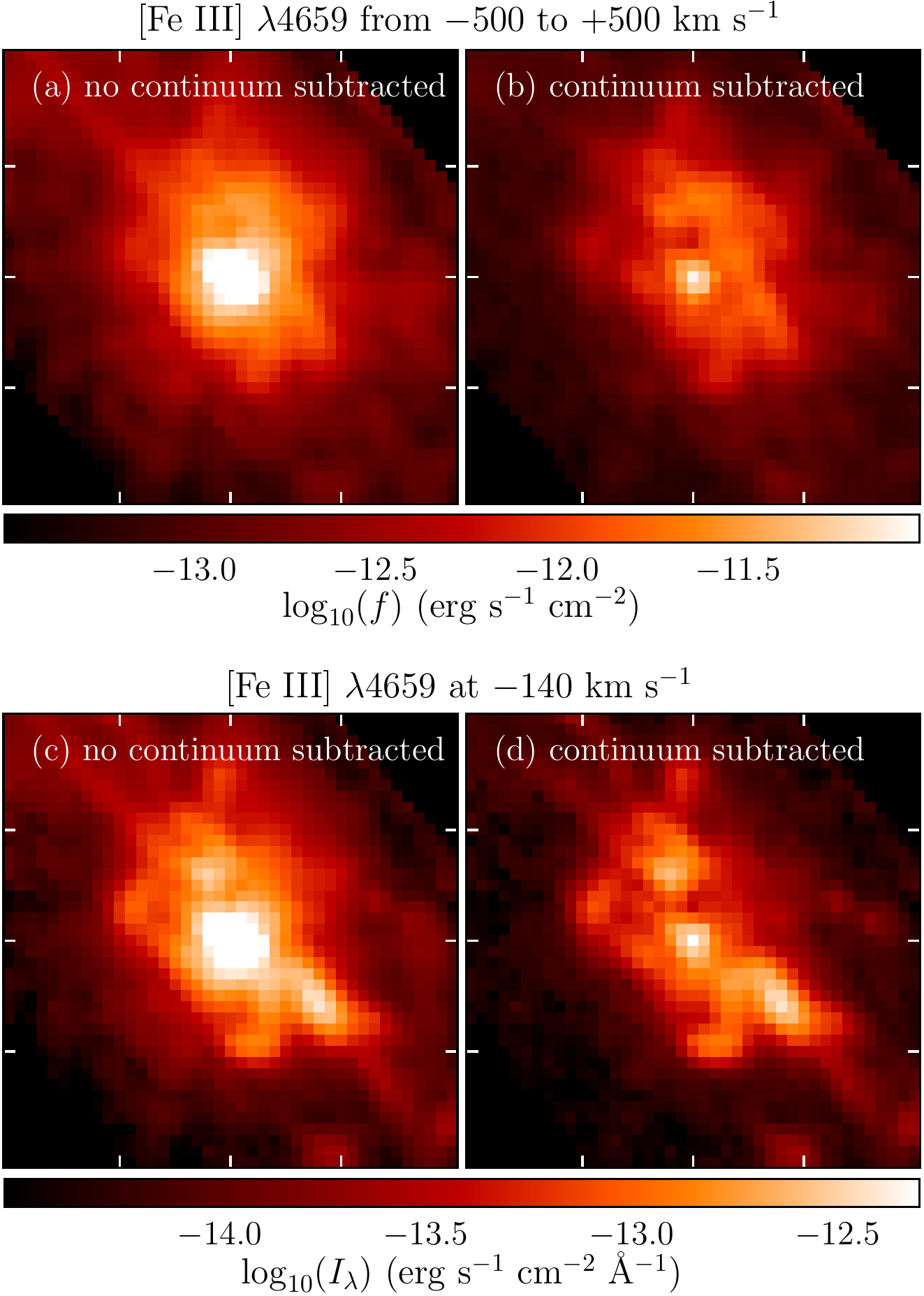}
\caption{{\bf Two examples of continuum subtraction for [Fe~III] 4659\AA\ recorded during the mapping visit near apastron, $\phi=$12.516.    a:} Image slice, 1000 \kms\  wide, which includes continuum. {\bf b:} Same image slice with continuum subtracted. {\bf c:} Image slice 40 \kms\ wide centered at $-$140 \kms\ including continuum. {\bf d:} Same image slice with continuum subtracted.  The dynamic range, as demonstrated by the color bar at the bottom of each figure, is set to maximize the information content. The angular size of all of these images is 2\arcsec $\times$ 2\arcsec. The velocity interval, 40 \kms, is comparable to  the spectrographic resolving power, $R=\lambda/\delta\lambda = $8000 or  33 \kms in velocity.\label{Subdemo}}
\end{figure}

Continuum emission is the major contribution to  spectra recorded when the \stis\ aperture is centered on, or near, \ecar\ and  is also a very significant component due to scattered starlight in the  surrounding nebular structures. The continuum flux must be subtracted in order to isolate emission from the extended wind-wind structures. Indeed,  the spectral region from 4563 to 4749\AA, recorded with the G430M grating set at the central wavelength, 4706\AA, has very few spectral intervals free from contributions from broadened stellar lines and narrow nebular lines. While many procedures were tested to subtract the continuum, the more successful approach was a multivariate fit across the recorded spectral region. By comparison, \cite{martin06}, \cite{mehner15} and \cite{Teodoro16} referenced measurements to a short interval at 4741 to 4743\AA\ and  4595 to 4605\AA. Test subtractions confirmed that the spectral interval around 4600\AA\ is contaminated by spatially resolved emission and absorption structures, and hence is unsuitable for continuum reference.

Our objective was to obtain velocity information for selected spectral lines. We found the best approach was to construct data cubes, with coordinates of right ascension, declination and velocity, from the mapping spectra with appropriate subtraction of continuum. Examples of image slices before and after continuum subtraction are presented in Fig. \ref{Subdemo}.  Each longslit spectrum was reduced by standard \stis\ data reduction tools. A routine, written using Interactive Data Language ({\bf IDL}\footnote{{\bf IDL} is a trademark of Exelis Visual Information Solutions, Inc,; http://www.exelisvis.com/ProductServices/IDL.aspx }), provided  continuum subtraction from each spectrum and resampled  the multiple spectra into 3-D data cubes with velocity  referenced to the transition wavelength: in this case, the [Fe~III] 4659.35\AA\ and the [Fe~II] 4815.88\AA\ lines. 

Subtraction of the continuum and careful selection of the velocity interval brings out considerable detail on the fossil wind structures as demonstrated in Fig. \ref{Subdemo}.  Images displayed in \ref{Subdemo}.a and .b extend from $-$500 to $+$500 \kms\ centered on [Fe~III] 4659.35\AA. In Fig. \ref{Subdemo}.a., the stellar core is surrounded by a diffuse haze and a hint of structure. Within each individual long-aperture, spatially resolved spectrum, a spectrum was extracted for each spatial element along the aperture. Then a slowly varying function was fitted to the spectrum to obtain the continuum level which was then subtracted.  Considerably more spatial structure then appears as shown in Fig. \ref{Subdemo}.b but which still spans the $-$500 to $+$500 \kms\ range.  Narrowing the velocity range to 40 \kms, slightly more than the 33 \kms\ resolution of the spectroscopic setup, reveals considerable differences in structures. In Fig. \ref{Subdemo}.c, which has no continuum subtracted, the stellar core dominates the image, but in Fig. \ref{Subdemo}.d, continuum has been subtracted which greatly suppresses the stellar core. Much more structural detail can be traced. 

All images in Figs. \ref{a-40} through \ref{a+220}, \ref{FeIIImosaic} and \ref{FeIImosaic}  have been continuum subtracted and normalized to the brightest structures. 
All images have a dynamic range from log$_{10}$ (4 $\times$ 10$^{-16}$ to 4 $\times$ 10$^{-13}$erg cm$^{-2} $sec$^{-1}$).

\section {The global changes of [Fe~III] and [Fe~II] across one binary period}
\label{global}

The velocity range, intervals and central velocities were chosen based upon observed structures. As demonstrated by \cite{gull09, Gull11}, the emission line structures are visible throughout the velocity range between $-$470 to $+$470 \kms, consistent with the models discussed by  \cite{Groh12}. 
We find no evidence of strong emission in the [Fe~III] or [Fe~II] lines blue-ward beyond the $-$460 \kms\ band (range  from $-$480 to $-$440 \kms). 
We chose to display the velocity images in 40 \kms\ intervals as that showed sufficient changes in structure in a finite number of frames.  While \cite{smith04_vel} measured the velocity range of H$_2$ in the Homunculus and estimated the system velocity, $V_{sys}=-$8.1$\pm 1$ \kms, we note that Weigelt D, as analyzed by \cite{zethson01, Zethson12} is centered at $-$45 \kms. Integrated profiles of [Fe~III] and [Fe~II] from the  2\arcsec $\times$ 2\arcsec\ maps produce narrow line components centered at $-$40 \kms with half-intensity levels at  $-$80 and $+$20 \kms. 

As we desired to track the slowly moving gas relative to the fossil wind structures, for the velocity range displays in Figs. \ref{FeIIImosaic} and \ref{FeIImosaic}, we displayed images centered at $-$60 and $-$20 \kms, on both sides of the slowly moving gas and extended the display range from $-$500 to $+$500 \kms. 

Instrumental and telescope diffraction effects minimally affect the velocity frames displayed in Figs. \ref{FeIIImosaic} and \ref{FeIImosaic}. The measured full width at half maximum width of the core component of \ecar\ averaged 0\farcs13 with small variations between visits.
These critically-phased observations were recorded at arbitrary PAs, dictated by \hst\ solar panel orientations. While a noticeable charge-transfer-inefficiency (CTI) trail, due to the aging CCD, and a diffraction ring are present in the reduced, broad-band integrated \stis\ imagery, continuum subtraction cancels out these contributions. Some hot column effects remain in the data, most noticeably in [Fe~II] +420 \kms\ iso-velocity frames (Fig. \ref{FeIImosaic}). Despite large changes in PA from visit to visit, the [Fe~III] and [Fe~II] structures persist, changing very slowly, corresponding to changes in expanding physical structures at their photoionization and critical densities.

\clearpage
\begin{landscape}
\begin{figure}

\centering
\includegraphics[width=9.2in,angle=0]{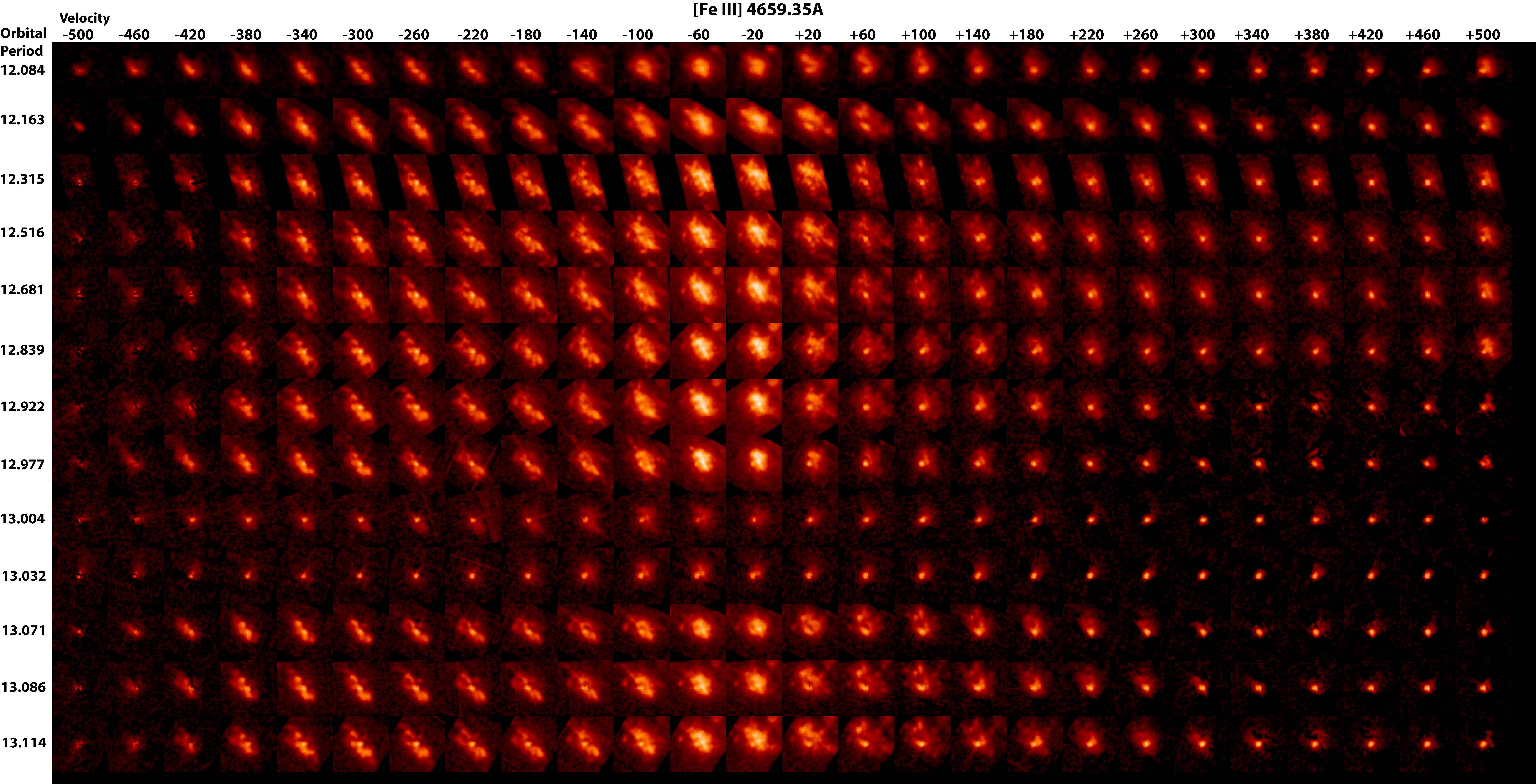}

\caption{{\bf Evolution of  highly ionized structures across the long high-ionization state through the low-ionization state and recovery: [Fe~III] 4659.35\AA\ mappings from phases $\phi=$ 12.084 to 13.114.} The [Fe~III] emission traces relatively dense (n$_e \approx\ $10$^7$cm$^{-3}$), gaseous regions that are directly photo-ionized by $>$16.2 eV FUV radiation. The initial mapping (top row $\phi=$ 12.084) shows the system in recovery from the low-ionization state induced by the periastron passage of \ecar~B, the UV source, deep within the extended wind of \ecar~A. The high-ionization state is well developed by $\phi=$ 12.163 and stays bright through 12.839, then rapidly fades to a stellar-like core by $\phi=$ 13.004, a few days after the onset of the low-ionization state. By $\phi=$ 13.071, recovery towards the high-ionization state is occurring. The extended complex at $-$60 and $-$20 \kms, which is the diffuse structure associated with the evolved Weigelt objects, develops a hook-shaped structure to the north and west of the central object. Note the lack of bright clumps in the $+$20 and $+$60 \kms\ velocity columns. Two major effects occur across the 5.54-year period: 1) expansion of arcs radially outward from \ecar, and 2) clockwise-tangential shifts of ionized regions due to movement of \ecar~B within the cavity and intervening absorbing material.
Detailed explanations of these many structures must await 3-D hydrodynamic modeling with radiative transfer. This is a display of the central 2\arcsec $\times$ 2\arcsec\ region around \ecar. Each continuum-subtracted frame, centered on \ecar, is an iso-velocity slice 40 \kms\  wide, displayed with the range from log$_{10}(4\times10^{-16}$) to  log$_{10}$(4$\times$10$^{-13}$) erg sec$^{-1}$ cm$^{-2}$. Each row displays velocity slices for a specific orbital phase recorded from June 2009 ($\phi=$ 12.084) to March 2015 ($\phi=$ 13.114) and each column displays velocity slices ranging from $-$500 to $+$500 \kms at increments of 40 \kms.  \label{FeIIImosaic}}

\centering
\end{figure}

\clearpage

\begin{figure}
\centering
\includegraphics[width=9.2in]{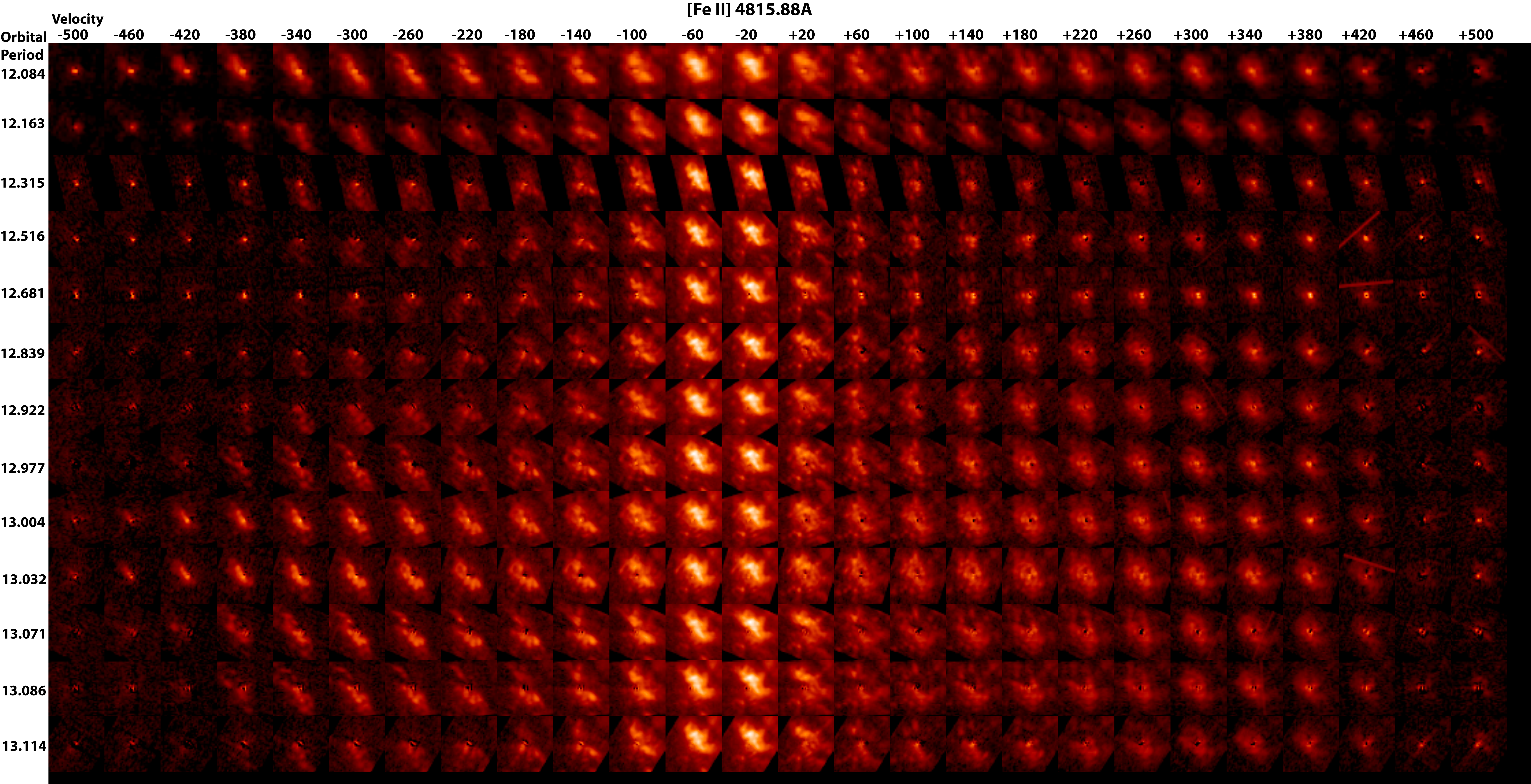}

\caption{{\bf Evolution of low-ionized structures across the long high-ionization state through the low-ionization state and recovery: [Fe~II] 4815.88\AA\ mappings  from phase $\phi=$ 12.084 to 13.114.} The  [Fe~II] traces regions where n$_e=2\times10^6$cm$^{-3}$, lower than traced by the [Fe~III] emission,  and  where iron is photo-ionized to Fe$^{+}$ by FUV radiation with energies  exceeding 7.9 eV. Throughout the 5.54-year period, the slowly moving gas ($-$60 and $-$20 \kms) is well defined as a hook-shaped structure to the north and west of \ecar. At $\phi=$ 12.084 (first row) relatively bright structures are present from $-$460 to $-$420 \kms. By $\phi=$ 12.163, the high velocity components are fading and continue to fade especially at the blue-shifted velocities, until 12.977. The [Fe~II] line is strongest across $\phi=$ 13.004 to 12.032, then begins to fade by $\phi=$ 13.071.  Especially at red-shifted velocities, many clumps appear in [Fe~II] extending out to the edges of the 2\arcsec $\times$ 2\arcsec frame. Likely these are clumps that are photo-ionized to Fe$^{++}$ across the high-ionization state, but have densities, n$_e<$ 10$^7$ cm$^{-3}$, but close to n$_e=$ 2$\times$10$^6$ cm$^{-3}$. Image display parameters are the same as in Fig. \ref{FeIIImosaic}. \label{FeIImosaic}}
\end{figure}

\end{landscape}
The rows of the two mosaics are paired by binary phase for comparison. The mappings began in June 2009 at $\phi=$ 12.084, as the system was recovering from the 2009.1 periastron event and ended 5.7 years later in March 2015 at $\phi=$ 13.114, just over one cycle later. Two mappings are very close to one cycle separation: $\phi=$ 12.084 and 13.086. A second near-pairing was accomplished with $\phi=$ 12.163 and 13.114. However both the $\phi=$ 12.084 and 12.163 mappings were accomplished with 0\farcs1 sampling, not the optimal sampling of 0\farcs05. Hence comparisons are  limited due to non-optimal sampling in the first two visits. Mappings were done at regular intervals across the long, high-ionization state, where spectrophotometry over past cycles showed relatively little change and, at critical phases, based upon predictions from the 3-D hydrodynamical models \citep{Madura12, madura13}.

The two lines selected for this study are relatively clear of weaker emission lines and broad stellar absorption features (Appendix \ref{reduce}). The [Fe~II] velocity range  from $-$500 to $+$500 \kms\ has no contamination from weak nebular emission lines, identified in spectra of Weigelt D \citep{Zethson12}. However the [Fe~III] velocity range has three weak lines that contribute flux at velocities beyond $+$340 \kms. While weaker relative to the [Fe~III] line by an order of magnitude, an [Fe~II]  line at $+$364 \kms\ and a (permitted) Fe line at $+$412 \kms\ remain across the low-ionization state when [Fe~III] disappears. As described in Appendix \ref{global}, the [Fe~II] 4815.88\AA\ emission line in the form of a data cube was shifted in velocity and adjusted in flux to null out contributions from these two lines in Fig. \ref{FeIIImosaic}. Hence while the mosaics are suitable for comparison at all velocities, the reader must be aware that a qualitative correction has been made in the [Fe~III] velocity frames at velocities red-shifted beyond $+$340 \kms.  

Complex spatial structures in [Fe~III] are visible across the long, high-ionization state  (Fig. \ref{FeIIImosaic}, $\phi=$ 12.084 to 12.977 and 13.071 to 13.114) that disappear during the relatively short, low-ionization state ($\phi=$ 13.004 to sometime beyond 13.032). 

While [Fe~II] is bright across the entire orbital period in the velocity bands centered on $-$60 and $-$20 \kms (Fig. \ref{FeIImosaic}), the blue-shifted and red-shifted velocity bands are brightest in [Fe~II] across the low-ionization state and fade considerably across the high-ionization state (a more detailed discussion on changes in the line fluxes of [Fe~III] and [Fe~II] is presented in Appendix \ref{flx}).

Many large-scale changes  are readily identifiable as due to photo-ionization or expansion.

\subsection{Changes due to photoionization}

\begin{enumerate}
\item  The [Fe~III] flux fades across the low-ionization state (Fig. \ref{FeIIImosaic}, $\phi=$ 13.004 to 13.032). Both lines are brightest in the $-$60 and $-$20 \kms\ velocity slices, which bound the $-$45 \kms\ peak velocity of the Weigelt objects \citep{Zethson12} and related debris associable with ejecta from the 1890s outburst \citep{weigelt12}.
\item As  [Fe~III] disappears, these structures reappear in [Fe~II] (Fig. \ref{FeIImosaic}, $\phi=$ 13.004 to 12.032).
\item Irregular [Fe~III] structures to the northeast and north that come and go. Many shift with time in a clockwise direction (Fig. \ref{FeIIImosaic}, $-$140 to $+$20 \kms).
\item Long spikes of [Fe~III]  extending to the northeast and southwest appear and disappear at $-$140 \kms\ as \ecar\ enters and leaves the low-ionization state. These spikes bound [Fe~II] emission at several phases (Fig. \ref{FeIIImosaic}, $\phi=$ 12.163, 12.977, 13.073 to 13.114).
\item  Numerous [Fe~II] clumps appear at red-shifted velocities exceeding +60 \kms. Most of these clumps are 0\farcs5 radially distant from \ecar.
\item The outermost structures disappear across the 5.7-year interval of observations likely due to photo-ionization effects.

\item Both lines are brightest in the $-$60 and $-$20 \kms\ velocity slices, which bound the $-$45 \kms\ velocity of the Weigelt objects \citep{Zethson12} and related debris associable with the ejecta from the 1890s outburst. However the [Fe~III] emission shrinks to the stellar position at all velocities across the 2014.6 periastron event at $\phi=$ 13.004 and 13.032. A very faint, diffuse halo persists extending  0\farcs4 to the northwest. By contrast the [Fe~II] strengthens at all velocities across the periastron event.
\end{enumerate}
\subsection{Changes due to expansion}

\begin{enumerate}
\item The [Fe~III] arcuate structures, displaced about 0\farcs4 to the northwest and south of \ecar, appear by $\phi=$ 12.163 in the $-$60 to $-$300 \kms\ velocity frames and expand radially through $\phi=$ 12.977 (Fig. \ref{FeIIImosaic}). These same arcs appear in [Fe~II] by $\phi=$ 12.922, brighten and persist across the low-ionization state, then fade by $\phi=$ 13.071.
\item A single [Fe~III] arc, displaced about 0\farcs2 to the northeast, appears by $\phi=$ 12.163 in the $+$140 to $+$340 \kms\ iso-velocity frames and expands until fading by $\phi=$ 12.977 (Fig. \ref{FeIIImosaic}). This arc also appears faintly in [Fe~II] across the low-ionization state (Fig. \ref{FeIImosaic}).
\item Multiple pairs of arcs appear in [Fe~II] to the west and south by $\phi=$ 12.315 and are traced through $\phi=$ 13.071. These pairs of arcs, described by \cite{Teodoro13}, expand radially from the \ecar\ position at initial separations in $\phi=$ 12.315 of 0\farcs2 and 0\farcs45. When modeled as shells, the characteristic expansion velocity of these red-shifted shells is about 470 \kms, somewhat larger than the terminal velocity, \vinf$_{, A}=$ 420 \kms\ determined by \cite{Groh12}.
\end{enumerate}

\section{The changing fluxes of [Fe~III] 4659 and [Fe~II] 4815\AA}\label{flx}

\cite{damineli08_period} measured the binary period using X-ray, optical and near infrared observations  to be 2022.74$\pm$1.3 days, referencing the zero phase of cycle 11 to JD 2452819.8 based upon the periodic disappearance of the narrow component to the He~I 6678\AA\ line. The 2014.6 drop in He~I 6678\AA\ was forecast to occur on JD 2456865.2. \cite{Teodoro16} modeled the behavior of  He~II 4686\AA\ across the three most recent spectroscopic minima, and determined that periastron occurred on JD 2456874.4$\pm$1.3 days, or 9.2 days after the disappearance of the narrow component. \cite{Teodoro16} found the best fit to the period to be 2022.7$\pm$0.3 days.
In this discussion to be consistent with published phase convention, we choose to use the zero phase, as defined by \cite{damineli08_period} namely, $\phi\ =\ $.0000 is when the narrow component of the He~I 6678\AA\ disappears. However we keep in mind that periastron is now thought to occur 9.2 days, or $\delta\phi\ =\ $0.0045 later. Modeling of the X-ray hardness ratio  also agrees with periastron being about nine days after disappearance of the He~I narrow component (Corcoran et al. in prep).

\begin{figure}
\centering
\includegraphics[width=\columnwidth]{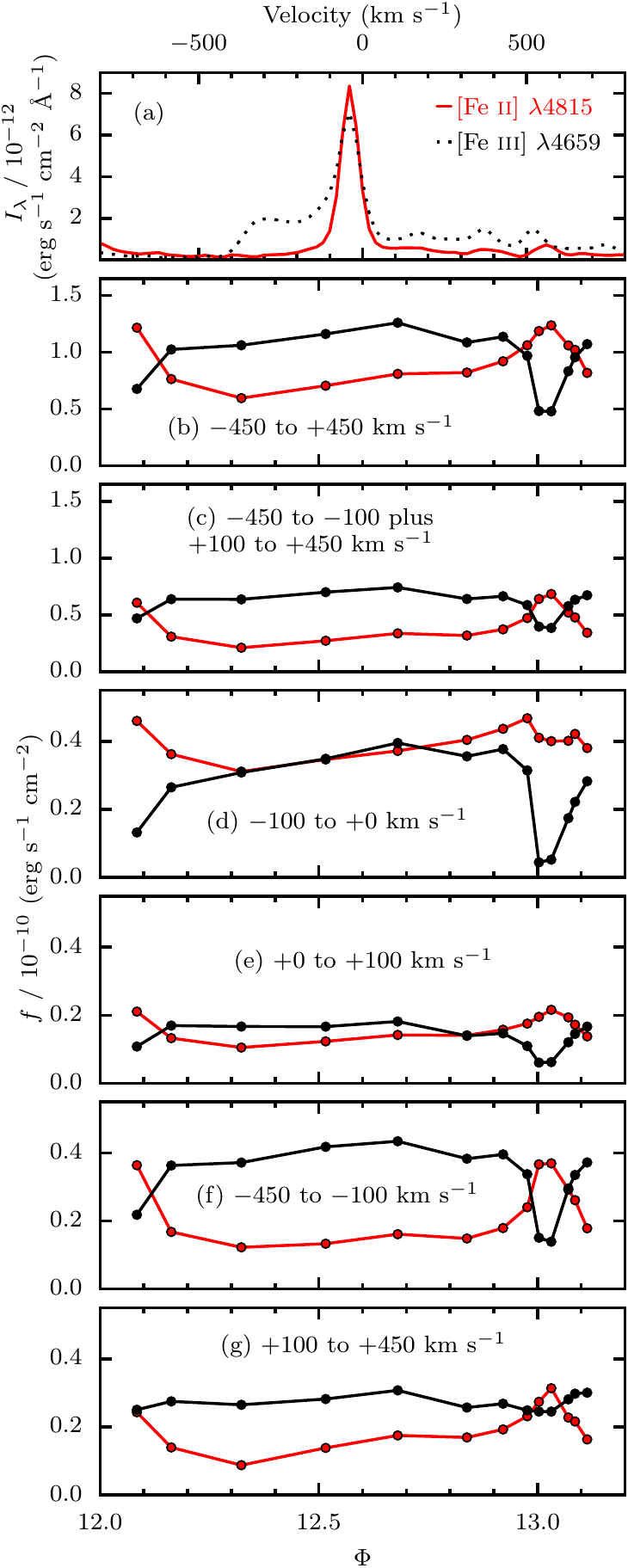}
\caption{{\bf Flux variations of [Fe~III] 4659.35\AA\ and [Fe~II] 4815\AA\ with binary phase. a:} Velocity line profiles. {\bf b:$-$ g:} Integrated fluxes  versus orbital phase. {\bf b:} Total line flux from $-$450 to $+$450 \kms. {\bf c:} Broad component excluding $-$100 to $+$100 kms, {\bf d:} $-$100 to 0 \kms\ component, which includes the Weigelt objects, {\bf e:} 0 to $+$100 component, {\bf f:} $-$450 to $-$100 \kms\ component, {\bf g:} $+$100 to $+$450 \kms\ component.}
\label{total}
\end{figure}

\begin{figure}
\includegraphics[width=\columnwidth]{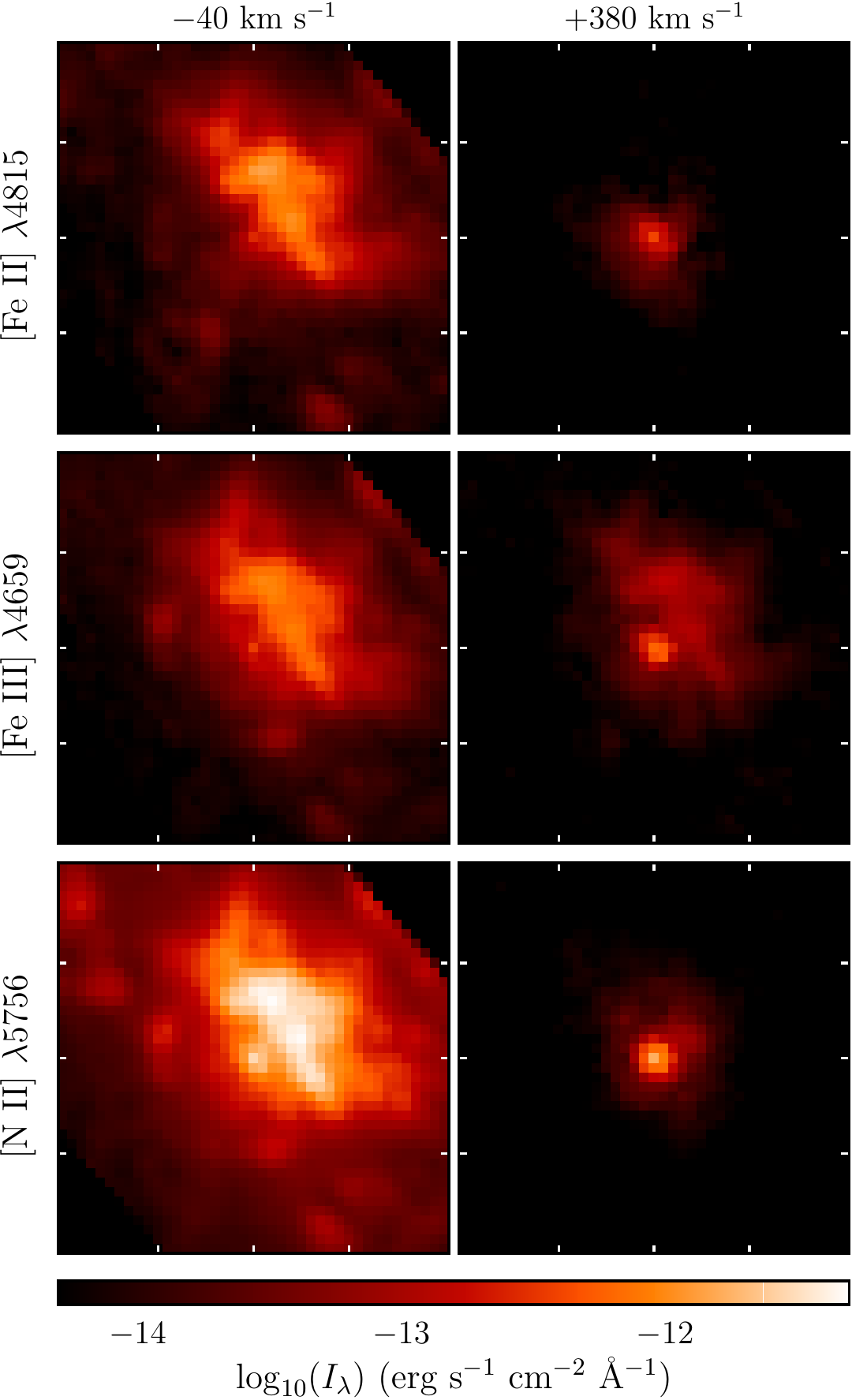}
\caption{{\bf Demonstration that the $+$380 \kms\ [Fe~III] images are contaminated by weak [Fe~II] and Fe~II line emission. Top row:} [Fe~II] 4815\AA. {\bf Middle row:}  [Fe~III] 4659\AA. {\bf Bottom row:} [N~II] 5756\AA.  Note that the hook-shaped structure appears in all three  images  centered at -40 \kms\ (left column), and faintly in the [Fe~III] $+$380 \kms\ image. However the hook is not present in the [Fe~II] and [N~II] $+$380 \kms\ images. Likewise the hook is not present in an  [Ar~III] $+$380 image (not shown) taken during the same mapping visit. Since the ionization potential, IP, of N$^+$ and Ar$^{++}$ are below and above the IP of Fe$^{++}$, the structure seen in the [Fe~III] $+$380 \kms\ image must be contamination by two weak lines:  Fe~II 4665.01\AA\ and [Fe~II] 4665.75\AA.  All  iso-velocity, 40 \kms-wide images are from mappings recorded in June, 2014 at phase, $\phi=$12.922. Mappings in [N~II]  required  a separate grating setting, 5734\AA,  and hence  an additional CVZ orbit. As a result, [N~II] mappings were done only at selected phases when deemed scientifically necessary. \label{compare}}
\end{figure}

 Consistent with the behavior of  He~I and highly ionized species \citep{damineli08_period}, we use the definition of  the high-ionization state to be when  both strong broad and narrow components of [Fe~III] are present, which lasts for about 5.2 years of the 5.54-year period (see Fig. \ref{total}). The low-ionization state, defined by   the interval when the narrow component of the high-ionization lines is absent, persists for   several months. However, a weak, broad component of  the high-ionization species, including Fe$^{++}$, remains. The broad, weak component is isolated to the stellar core and a very diffuse region extending to the northwest as seen in the image slices across the low-ionization state in Fig. \ref{FeIIImosaic}. 
 
 From our mappings, measures of these components are plotted in Fig. \ref{total}. The flux distributions with velocity at apastron are plotted for the [Fe~III] 4659\AA\ and the [Fe~II] 4815\AA\  in Fig. \ref{total}.a. 
 
 All other subplots in Fig. \ref{total}\  display integrated fluxes plotted against orbital phase. Changes are greatest across the low-ionization state. In Fig. \ref{total}.b, [Fe~II] (red) drops as [Fe~III] (black) increases with the onset of the high-ionization state (12.084 to 12.163 and 13.071 to 13.114), and [Fe~II]  increases as [Fe~III] fades as the low-ionization state approaches (12.922 to 13.004). Separation   into  a total broad components, $-$450 to $-$100 \kms plus $+$100 to $+$450 \kms,  leads to  less pronounced changes (Figs.  \ref{total}.c).
The narrow, blue-shifted components, $ -$100 to 0 \kms\ (Fig. \ref{total}.d) , show the largest changes across the low-ionization state. The narrow, red-shifted components, 0 to $+$100 \kms, follow  much weaker,  changes (Fig. \ref{total}.e). 
 
\begin{figure*}
\includegraphics[width=7 in]{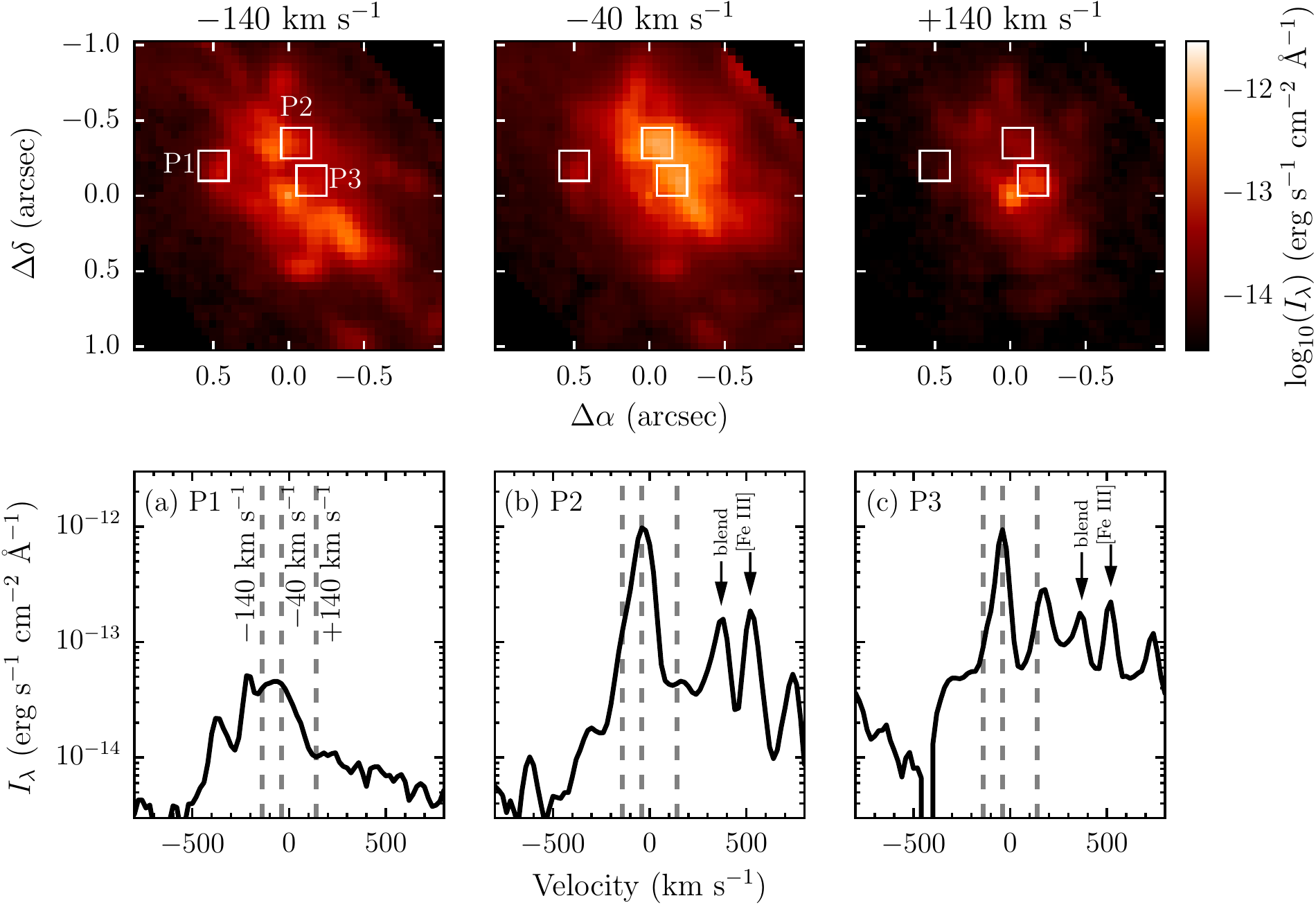}
\caption{{\bf Sample spectra from selected portions of the interacting winds. Top row:} Image slices 40 \kms\ wide centered at $-$140, $-$40 and $+$140 \kms. {\bf Bottom row:} Extracted spectra from  0\farcs1
$ \times$ 0\farcs1 areas. Left: Spectrum extracted near the peak of the $-$140 \kms\ component, P1, to the west of \ecar. Center: Spectrum extracted at the peak of emission, P2,  associable with Weigelt D. Right: Spectrum extracted at the peak of a red-shifted [Fe~II] arc, P3. The data presented in this figure were recorded November 20, 2011 at \ecar's orbital phase, $\phi=$12.516.\label{grid}}
\end{figure*}

 The blue-shifted, broad component, $-$450 to $-$100 \kms\ (Fig. \ref{total}.f), shows a pronounced contrast between high-low-ionization state and low-ionization states, with the [Fe~II] line being very faint during the high-ionization state. The change from high- to low-ionization state is much less pronounced for the red-shifted broad component, $+$100 to $+$450 \kms\ (Fig. \ref{total}.g), partially due to the weak [Fe~II] and Fe~II emission-line contamination contributed to [Fe~III]  in the $+$300 to $+$450 \kms\ velocity interval (see Fig. \ref{grid}). 
 
We demonstrate that the red-shifted component of [Fe~III] is contaminated  in two ways:
i) structures seen at $+$380 \kms\ for the [Fe~II], [Fe~III] and [N~II] lines (see Section \ref{IP}) and ii) spectra of selected regions as seen in [Fe ~III] (Section \ref{spec}).

\subsection {Changes in structure by ionization potential, IP}
\label{IP}
Velocity slices at $-$40 and $+$380 \kms\ are presented in Fig. \ref{compare} for [Fe~II] (.a), [Fe~III] (.b) and [N~II] (.c). The hook structure noted in Section \ref{slow} and Figs. \ref{field} and \ref{a-40} is always present in $-$40 \kms\ images of [Fe~II], but disappears across the low-ionization state in [Fe~III]. Indeed across the low-ionization state, the broad component of [Fe~III] is in the stellar core region and extends faintly to the northwest. 

Comparison to the [Fe~II] and [N~II] velocity slices across the velocity range shows that [Fe~II] strengthens across the low-ionization state at red-shifted velocities, but [N~II] virtually disappears. At $+$380 \kms\, only a faint core emission exists at the stellar core position. Yet  a hook-shaped structure persists in the [Fe~III] at this velocity, peaking at about $+$380 \kms. Likewise, no hook-shaped emission is present in the several [Ar~III] 7137\AA\ maps recorded at selected times during the orbital cycle. Since the hook-shaped structure does not exist for N$^+$ nor Ar$^{++}$ emissions with IP= 14.5, 27.6 eV, then the hook-shaped structures must be attributed to the [Fe~II] 4665.75\AA\ and Fe 4665.01\AA\ lines.

\subsection {Velocity variations of [Fe~III] for selected positions}
\label{spec}
The forbidden emission varies considerably from point to point across the fossil winds. We demonstrate this for [Fe~III] near apastron in Fig. \ref{grid} by presenting three spectral extractions from representative locations:  P1, in  the direction of a blue-shifted arcs expanding to the west-northwest, P2, to the north-northeast at the position of Weigelt D, a slowly moving emission clump thought to have originated during the 1890s event, and P3, a red-shifted clump almost due east of \ecar.

The fluxes for each spectrum are plotted on the same intensity scale. Likewise, the image slices are displayed on the same log scale. The narrow velocity component, $-$40 \kms, is brightest from the slowly moving gas, which includes Weigelt D. Yet weaker emission extends at red-shifted velocities that are not of [Fe~III] origin. The peak near $+$375 \kms\  originates from a blend of weak [Fe~II] 4665.75\AA\ and Fe~II  4665.01\AA\ and the peak at 510 \kms\ is due to a weaker [Fe~III] 4668.06\AA\  \citep{Zethson12}. The blend at $+$375 \kms\ changes with orbital phase similar to the [Fe~II] 4815.88\AA\ and the $+$510 \kms\ emission appears and disappears similar to the [Fe~III] 4659.35\AA\ emission. Note that in Fig. \ref{FeIIImosaic} a flux-adjusted, velocity-shifted blend of the [Fe~II] 4815.88\AA\ emission was subtracted to minimize its presence in the [Fe~III] emission. The [Fe~III] 4668.01\AA\ emission was not subtracted, so it appears in the $+$500 \kms\ velocity images. 

We examined the spatial-velocity cubes of [N~II] and [Ar~III] recorded at selected times during this program. Both have no identified narrow-line emission in the 0 to $+$600 \kms\ velocity ranges \citep{Zethson12}. Neither emission line shows the hook-shaped structure. Indeed the [N~II] in the red-shifted velocity frames show a converging shell that disappears beyond $+$420 \kms. Hence the subtraction of the [Fe~II] from the [Fe~III] mosaic (Fig. \ref{FeIIImosaic}) appears to be reasonably accurate and successful.

At P1, the $-$40 \kms\ emission has dropped twenty-fold, and a comparable flux is present extending to $-$200 \kms\ (see Fig. \ref{a-140}). Red-shifted emission is considerably weaker. 

At P3, while the $-$40 \kms\ narrow emission is strongest, a narrow-line emission pops up at $+$200 \kms. The structural extent of this $+$200 \kms\ emission is very different from the $-$40 \kms\ emission (see Fig. \ref{a+220}).

\bsp	
\label{lastpage}

\end{document}